\documentclass[preprint]{aastex}

\received{}
\accepted{}

\shortauthors{Richards et al.}
\shorttitle{SDSS Quasar Photometry}

\title{Colors of 2625 Quasars at $0<z<5$ Measured in the Sloan Digital
Sky Survey Photometric System\footnote{Based on observations obtained
with the Sloan Digital Sky Survey; with the Apache Point Observatory
3.5-meter telescope, which is owned and operated by the Astrophysics
Research Consortium; with the Hobby-Eberly Telescope, which is a joint
project of the University of Texas at Austin, the Pennsylvania State
University, Stanford University, Ludwig-Maximillians-Universit\"at
M\"unchen, and Georg-August-Universit\"at G\"ottingen; with the
2.1-meter telescope at Kitt Peak National Observatory; and at the
W. M. Keck Observatory, which is operated as a scientific partnership
among the California Institute of Technology, the University of
California, and NASA, and was made possible by the generous financial
support of the W. M. Keck Foundation.}}

\author{
Gordon T. Richards\altaffilmark{2,3},
Xiaohui Fan\altaffilmark{4,5},
Donald P. Schneider\altaffilmark{2},
Daniel E. Vanden Berk\altaffilmark{6},
Michael A. Strauss\altaffilmark{4},
Donald G. York\altaffilmark{3,7},
John E. Anderson, Jr.\altaffilmark{6},
Scott F. Anderson\altaffilmark{8},
James Annis\altaffilmark{6},
Neta A. Bahcall\altaffilmark{4},
Mariangela Bernardi\altaffilmark{3},
John W. Briggs\altaffilmark{3},
J. Brinkmann\altaffilmark{9},
Robert Brunner\altaffilmark{10},
Scott Burles\altaffilmark{3,6},
Larry Carey\altaffilmark{8},
Francisco J. Castander\altaffilmark{3,11},
A. J. Connolly\altaffilmark{12},
J. H. Crocker\altaffilmark{13},
Istv\'an Csabai\altaffilmark{13,14},
Mamoru Doi\altaffilmark{15},
Douglas Finkbeiner\altaffilmark{16},
Scott D. Friedman\altaffilmark{13},
Joshua A. Frieman\altaffilmark{3,6},
Masataka Fukugita\altaffilmark{17},
James E. Gunn\altaffilmark{4},
Robert B. Hindsley\altaffilmark{18},
\v{Z}eljko Ivezi\'{c}\altaffilmark{4},
Stephen Kent\altaffilmark{3,6},
G. R. Knapp\altaffilmark{4},
D.Q. Lamb\altaffilmark{3},
R. French Leger\altaffilmark{8},
Daniel C. Long\altaffilmark{9},
Jon Loveday\altaffilmark{19},
Robert H. Lupton\altaffilmark{4},
Timothy A. McKay\altaffilmark{20},
Avery Meiksin\altaffilmark{21},
Aronne Merrelli\altaffilmark{10,22},
Jeffrey A. Munn\altaffilmark{23},
Heidi Jo Newberg\altaffilmark{24},
Matt Newcomb\altaffilmark{22},
R. C. Nichol\altaffilmark{22},
Russell Owen\altaffilmark{8},
Jeffrey R. Pier\altaffilmark{23},
Adrian Pope\altaffilmark{13,22},
Michael W. Richmond\altaffilmark{25},
Constance M. Rockosi\altaffilmark{3},
David J. Schlegel\altaffilmark{4},
Walter A. Siegmund\altaffilmark{8},
Stephen Smee\altaffilmark{13,26},
Yehuda Snir\altaffilmark{22},
Chris Stoughton\altaffilmark{6},
Christopher Stubbs\altaffilmark{8},
Mark SubbaRao\altaffilmark{3},
Alexander S. Szalay\altaffilmark{13},
Gyula P. Szokoly\altaffilmark{27},
Christy Tremonti\altaffilmark{13},
Alan Uomoto\altaffilmark{13},
Patrick Waddell\altaffilmark{8},
Brian Yanny\altaffilmark{6},
Wei Zheng\altaffilmark{13}
}

\altaffiltext{2}{Department of Astronomy and Astrophysics, The Pennsylvania State University, University Park, PA 16802}
\altaffiltext{3}{The University of Chicago, Department of Astronomy and Astrophysics, 5640 S. Ellis Ave., Chicago, IL 60637}
\altaffiltext{4}{Princeton University Observatory, Princeton, NJ 08544}
\altaffiltext{5}{Institute for Advanced Study, Olden Lane, Princeton, NJ 08540}
\altaffiltext{6}{Fermi National Accelerator Laboratory, P.O. Box 500, Batavia, IL 60510}
\altaffiltext{7}{The University of Chicago, Enrico Fermi Institute, 5640 S. Ellis Ave., Chicago, IL 60637}
\altaffiltext{8}{University of Washington, Department of Astronomy, Box 351580, Seattle, WA 98195}
\altaffiltext{9}{Apache Point Observatory, P.O. Box 59, Sunspot, NM 88349-0059}
\altaffiltext{10}{Department of Astronomy, California Institute of Technology, Pasadena, CA 91125}
\altaffiltext{11}{Observatoire Midi Pyrenees, 14 ave Edouard Belin, Toulouse, F-31400, France}
\altaffiltext{12}{Department of Physics and Astronomy, University of Pittsburgh, Pittsburgh, PA 15260}
\altaffiltext{13}{Department of Physics and Astronomy, The Johns Hopkins University, 3701 San Martin Drive, Baltimore, MD 21218}
\altaffiltext{14}{Department of Physics of Complex Systems, E\"otv\"os University, P\'azm\'any P\'eter s\'et\'any 1}
\altaffiltext{15}{Department of Astronomy and Research Center for the Early Universe, School of Science, University of Tokyo, Hongo, Bunkyo, Tokyo, 113-0033, Japan}
\altaffiltext{16}{University of California at Berkeley, Departments of Physics and Astronomy, 601 Campbell Hall, Berkeley, CA 94720}
\altaffiltext{17}{University of Tokyo, Institute for Cosmic Ray Reserach, Kashiwa, 2778582, Japan}
\altaffiltext{18}{Remote Sensing Division, Code 7215, Naval Research Laboratory, 4555 Overlook Ave. SW, Washington, DC 20375}
\altaffiltext{19}{Astronomy Centre, University of Sussex, Falmer, Brighton BN1 9QJ, United Kingdom}
\altaffiltext{20}{University of Michigan, Department of Physics,	500 East University, Ann Arbor, MI 48109}
\altaffiltext{21}{Royal Observatory, Edinburgh, EH9 3HJ, United Kingdom}
\altaffiltext{22}{Dept. of Physics, Carnegie Mellon University, 5000 Forbes Ave., Pittsburgh, PA-15232}
\altaffiltext{23}{U.S. Naval Observatory, Flagstaff Station, P.O. Box 1149, Flagstaff, AZ  86002-1149}
\altaffiltext{24}{Physics Department, Rensselaer Polytechnic Institute, SC1C25, Troy, NY 12180}
\altaffiltext{25}{Physics Department, Rochester Institute of Technology, 85 Lomb Memorial Drive, Rochester, NY 14623-5603}
\altaffiltext{26}{Department of Astronomy, University of Maryland, College Park, MD 20742-2421}
\altaffiltext{27}{Astrophysikalisches Institut Potsdam, An der Sternwarte 16, D-14482 Potsdam, Germany}

\begin{document}

\begin{abstract}

We present an empirical investigation of the colors of quasars in the
Sloan Digital Sky Survey (SDSS) photometric system.  The sample
studied includes 2625 quasars with SDSS photometry: 1759 quasars found
during SDSS spectroscopic commissioning and SDSS followup observations
on other telescopes, 50 matches to FIRST quasars, 573 matches to
quasars from the NASA Extragalactic Database, and 243 quasars from two
or more of these sources.  The quasars are distributed in a 2.5 degree
wide stripe centered on the Celestial Equator covering $\sim529$
square degrees.  Positions (accurate to $0.2\arcsec$) and SDSS
magnitudes are given for the 898 quasars known prior to SDSS
spectroscopic commissioning.  New SDSS quasars, which range in
brightness from $i^{*} = 15.39$ to the photometric magnitude limit of
the survey, represent an increase of over 200\% in the number of known
quasars in this area of the sky.

The ensemble average of the observed colors of quasars in the SDSS
passbands are well represented by a power-law continuum with
$\alpha_{\nu} = -0.5$ ($f_{\nu} \propto \nu^{\alpha}$) and are close
to those predicted by previous simulations.  However, the
contributions of the ``small blue (or $3000\,{\rm \AA}$) bump'' and
other strong emission lines have a significant effect upon the colors.
The color-redshift relation exhibits considerable structure, which may
be of use in determining photometric redshifts for quasars from their
colors alone.  The range of colors at a given redshift can generally
be accounted for by a range in the optical spectral index with a
distribution $\alpha_{\nu}=-0.5\pm0.65$ (95\%~confidence), but there
is a red tail in the distribution.  This tail may be a sign of
internal reddening, especially since fainter objects at a given
redshift tend to exhibit redder colors than the average.  Finally, we
show that there is a continuum of properties between quasars and
Seyfert galaxies and we test the validity of the traditional dividing
line ($M_B=-23$) between the two classes of AGN.

\end{abstract}


\section{Introduction}

Quasar identification in multicolor surveys relies on the fact that
the spectral energy distributions of stars and quasars produce
different colors in broad-band photometric systems.  At low redshift
($z\le2.2$), the lack of a Balmer jump in quasars separates them from
hot stars; at higher redshifts the presence of the strong
Lyman-$\alpha$ emission line and absorption by the Lyman-$\alpha$
forest cause the broad-band colors of quasars to become increasingly
redder with redshift \citep{san65,gsl86,whi+91,fan99}.  Historically,
large area surveys have been based on photographic plates; the limited
photometric accuracy of this technique ($\sim 0.1$ mag) has been one
of the primary limitations of the effectiveness of their quasar
selection.

The Sloan Digital Sky Survey (SDSS; \citealt{yor+00}) will survey
approximately one quarter of the Celestial Sphere in five broad bands
spanning the optical region of the spectrum.  The expected photometric
errors in each filter are less than 5\% for objects brighter than
$19^{\rm th}$ magnitude; this should allow identification of
approximately 100,000 quasars.  In this paper we study the photometric
properties in the SDSS system of over 2600 quasars in a stripe
$2.5^{\circ}$ wide centered on the Celestial Equator; the total area
covered is $\sim 529$ square degrees.  Over 1700 of these objects were
discovered in the past year and are examined here for the first time.

This paper acts as an empirical companion to \citet{fan99}, which
presented theoretical and simulated colors of quasars and other point
sources in the SDSS system.  Although the SDSS photometric system is
still uncertain at the few percent level, the photometry is better
than has ever been accomplished previously for a large area sample of
quasars.  With the SDSS photometry, we can detect features in the
color-redshift distribution that are smaller than the typical error in
the colors of quasars from photographic surveys.  Using this
information, we examine a number of long-standing issues in quasar
science.

Section~2 of the paper describes the observations and in Section~3 we
describe the catalogs that were used to construct the database for
this study.  In Section~4 we examine the observed colors of quasars
and compare them to the values expected from simulations of their
spectral energy distributions, and in Section~5 we discuss how these
results will impact a number of issues, including quasar target
selection in the SDSS.  Finally, Section~6 presents our conclusions.

\section{SDSS Imaging}

The Sloan Digital Sky Survey \citep{yor+00} will provide accurate
photometry and spectroscopy for over $10,000$ square degrees of sky.
The optical magnitudes of detected objects are measured nearly
simultaneously through five broad-band filters ($u'$, $g'$, $r'$,
$i'$, and $z'$; \citealt{fig+96}) with nominal effective wavelengths
of $3540\,{\rm \AA}$, $4760\,{\rm \AA}$, $6280\,{\rm \AA}$,
$7690\,{\rm \AA}$ and $9250\,{\rm \AA}$, complete to limiting point
source magnitudes of 22.3, 22.6, 22.7, 22.4 and 20.5, respectively
(signal-to-noise ratio 5:1).  Because the definition of the
photometric system is not yet finalized, measured magnitudes are
quoted using asterisks ($u^{*}$, $g^{*}$, $r^{*}$, $i^{*}$, and
$z^{*}$) to represent preliminary photometry; the filters themselves
are referred to as $u'$, $g'$, $r'$, $i'$, and $z'$.  SDSS magnitudes
are based on the AB magnitude system \citep{og83}, and are given as
asinh magnitudes \citep{lgs99}, which have nice properties at low flux
limits.  For objects fainter than the $5\sigma$ flux limit given
above, the asinh magnitude will deviate from normal magnitudes.  The
analysis in this paper should not be affected by these differences;
however, some of the high-redshift quasars have $u'$ and $g'$
magnitudes fainter than these limits.

The survey is being done with a dedicated 2.5m telescope
\citep{sie+00}.  The telescope has a wide, well-corrected field and is
equipped with a large mosaic CCD camera \citep{gcr+98} and a pair of
fiber-fed spectrographs \citep{uom+00a}.  The camera utilizes thirty
$2048\times2048$ CCDs which take the data in drift-scanning
(time-delay-and-integrate, or TDI) mode with a total integration time
of 54.1 seconds per filter.  The imaging data are obtained using the
data acquisition system \citep{pbm+94,ann+00} at the Apache Point
Observatory (APO) and are automatically processed through a set of
software pipelines \citep{ken+00}.  The photometric pipeline
\citep{lup+00} reduces the imaging data, measuring positions,
magnitudes, and shape parameters for all detected objects.  The
photometric pipeline uses information from the astrometric pipeline
\citep{pie+00} and the photometric calibration telescope (PT;
\citealt{smi+00,uom+00b,tuc+00}).  After final photometric calibration
of the data, the outputs, together with all the observing and
processing information, are loaded into the operational database
\citep{yan+00}.  The final parameters are stored in an object-oriented
searchable database (SX; \citealt{sza+00}).  For an explanation of
SDSS technical terms used in the text, please refer to \citet{yor+00}.

Preliminary analysis shows that there are differences between the real
SDSS transmission curves and those shown by \citet{fig+96}; see
\citet{fan+00}, Appendix A, for a discussion.  These differences are
believed to be due to the fact that the filters are in vacuo and are
not exposed to the air.  For a quasar with a typical power-law
spectrum of $f_{\nu}\propto\nu^{-0.5}$ we find that the effective
wavelengths of the actual 2.5m filters are $3651$, $4679$, $6175$,
$7494$, and $8873\,{\rm \AA}$ instead of $3544$, $4770$, $6231$,
$7625$ and $9135\,{\rm \AA}$, respectively for $u'$, $g'$, $r'$, $i'$,
and $z'$.  (Since the spectral energy distribution of a quasar is
different from the standard stars used to define the SDSS photometric
system, the latter effective wavelengths of the passbands are
different from those quoted earlier.)  These offsets are quite large,
but they produce only small effects on the measured colors of the
objects discussed herein ($\sim 0.005\,{\rm mag}$).  The most current
transmission curves are used in all of the analysis throughout the
paper.  More details on the transmission curves of the 2.5m camera
will be shown in a future paper \citep{doi+00}.

Data from the imaging camera on the 2.5-m are calibrated against
patches of the sky observed by the PT.  The PT also observes standard
stars in order to determine nightly extinction coefficients.  While
the quality of the current SDSS photometric data is more than adequate
for our purposes, the final calibration of the SDSS system is not yet
in place, due in part to the SDSS transmission curve uncertainties
discussed above.  However, the quality of the data in hand suggests
that any changes to the quasar color-color and color-redshifts
relations will be relatively minor, and should not have a significant
impact on the results presented herein.  

The SDSS data analyzed herein include the photometric catalog from the
four best equatorial scans that were taken between 1998 September 19
and 1999 March 22, which include data ``runs'' 94, 125, 752 and 756,
as described in Table~\ref{tab:tab1}.  These runs were observed and
processed as part of the commissioning phase of the SDSS; they were
acquired with the telescope parked on the meridian while pointed at
the Celestial Equator.  The density of previously known quasars is
highest on the Celestial Equator, making these data particularly
appropriate for this study.  Runs 94 and 125 are from the south
Galactic cap, whereas runs 752 and 756 are from the north Galactic
cap.  A fraction of the data from these runs have been re-reduced.
The mean difference between the old magnitudes and the new magnitudes
is (0.043,0.008,0.007,0.006,0.006) in ($u^*$, $g^*$, $r^*$, $i^*$,
$z^*$) coordinates.  Since these differences are negligible, we will
use the ``old'' magnitudes throughout for the sake of consistency.

\section{Quasar Catalog}

In order to study the colors of quasars in the SDSS photometric
system, we first generated a (full sky) master catalog of all known
quasars including the newly discovered quasars from the SDSS.  We then
matched the equatorial SDSS photometric catalog to the equatorial
regions of the master catalog.  The master quasar catalog used in this
study includes quasars from four distinct sources.  These four
subcatalogs of quasars were combined to form a larger combined
catalog.  The subcatalogs are described in the following sections.

\subsection{NED}

One third of the quasars studied herein are previously known quasars
that have been cataloged by the NASA Extragalactic Database
(NED)\footnote{The NASA/IPAC Extragalactic Database (NED) is operated
by the Jet Propulsion Laboratory, California Institute of Technology,
under contract with the National Aeronautics and Space
Administration.}.  Our NED subcatalog includes all of the 12,987
objects that NED classified as quasars with known redshifts as of 2000
June 22.  Many of these objects will also appear in the other
subcatalogs discussed later.

The NED catalog is a compilation of results from the literature.  Any
bias in the original selection of these objects will also be manifest
in this sample.  One of the most important selection effects is the
definition of what constitutes a quasar.  Typically, quasars are
defined to be active galactic nuclei (AGN) brighter than $M_B=-23$
($H_{\rm o} = 50\,{\rm km\,s^{-1}\,Mpc^{-1}}$, $q_{\rm o} = 0.5$);
however, Seyfert galaxies, the fainter cousins of quasars, are often
included in compilations of quasars.  Such objects from NED are
included in our combined catalog for sake of completeness even though
their colors may be contaminated by their host galaxy.  See
\S~\ref{sec:seyferts} for more on this subject.

The primary input to the NED quasar catalog is the catalog produced by
\citet[hereafter HB89] {hb89}, which includes over 3500 quasars.  As
with the NED catalog, HB89 is itself a compilation from the
literature.  The largest uniform sample of quasars ($\sim 1000$) in
NED comes from the Large Bright Quasar Survey (LBQS; \citealt{hfc95}).
Since LBQS quasars were selected from objective prism plates, the LBQS
catalog may include quasars with stronger than average emission lines.
Other optical quasar surveys that contribute significantly to the
catalog of known quasars include the AAT Survey \citep{bfs+90}, the
CFHT Survey \citep{cjd+88}, the Second Byurakan Survey (SBS;
\citealt{SLC+93}), and the Palomar Grism Surveys \citep{ssg94,ssg99}.
Not all of these surveys are color-selected surveys; each has its own
selection effects.  The AAT survey was an ultra-violet excess (UVX)
survey and is biased towards $z<2.2$.  The CFHT survey used a blue
``grens'' to do slitless spectroscopy and was sensitive to $z<3.3$.
The SBS survey was an objective prism survey, whereas the Palomar
Grism Surveys used a grism for candidate selection and slit
spectroscopy for followup observations.  Many of the radio-detected
quasars in NED come from \citet{vcv96}, which, like HB89, is a
compilation based on a variety of sources.  The brightest of these
objects are 3C sources \citep{sma+85}, which are brighter than
$9\,{\rm Jy}$ at $178\,{\rm MHz}$.

\subsection{FIRST}

We have attempted to recover, in the SDSS data, quasars detected by
the VLA FIRST Survey \citep{bwh95}, in particular those discovered by
the FIRST Bright Quasar Survey (FBQS; \citealt{gbw96,wbg+00,bec+01})
and the FIRST Faint Quasar Survey (FFQS; \citealt{blg98}).  Many of
the FBQS quasars also appear in the NED catalog; however, the FBQS is
an ongoing survey, so we have included these FIRST quasars explicitly.
Although the SDSS will be selecting FIRST sources as quasar
candidates, it is valuable to include in our sample those FIRST
objects that have already been confirmed as quasars.  As with the
radio-selected quasars from NED, FIRST quasars are valuable because
they have different selection biases than optically-selected quasars.
In particular, they can be used to determine the fraction of quasars
embedded in the stellar locus and to analyze the effects of reddening
and extinction upon quasar target selection.

\subsection{SDSS Spectroscopic Commissioning}

The largest part of the sample comes from spectroscopic confirmation
of quasar candidates from SDSS spectroscopic commissioning
\citep{uom+00a}.  The SDSS uses two double fiber-fed spectrographs;
each has separate red and blue cameras, with a dichroic splitting the
light at $\sim 6000\,$\AA.  The fibers are hand-plugged into an
aluminum plate (640 fibers per plate); the correspondence between hole
number and fiber number is determined by an automated plate mapper.
The entire fiber, plate and slit assembly is modular and is removed
and replaced for each spectroscopic exposure.

The fibers subtend $3\arcsec$ at the focal plane.  The resulting
spectra cover the range from 3800\AA\ to 9200\AA\ with a resolution of
roughly 1800, over 4096 pixels.  Typical exposure times are 45 minutes
to an hour, broken into single exposures of 15 minutes each.
Reduction on the mountain confirms that the signal-to-noise ratio of
the faintest objects meets survey specifications.  The wavelength
calibration is appreciably better than the specifications of 0.1\AA.
The flux calibration is currently only approximate, and is carried out
with observations of F subdwarfs.  The spectrographs are very high
throughput, approaching 30\% in the red; the resulting spectra have
quite high signal-to-noise ratio, of order 4 per pixel for objects as
faint as $r'=20$.

Quasar candidates for spectroscopy were selected by a preliminary
version of the SDSS Quasar Target Selection Algorithm
\citep{new+00,ric+01}, which is primarily a color-selection algorithm.
The primary selection criterion is that objects be outliers from the
stellar locus, which means that quasars with unusual colors will be
targeted.  Since one of the goals of the commissioning phase of this
project is to define this algorithm, the quasars selected with the
current algorithm cannot be considered representative of the final
SDSS quasar sample, or in any way to constitute a complete sample.  In
particular, these are color selected objects selected with color cuts
that are not yet final (moreover the color cuts have changed with
time), and the quasars do not cover a uniform area on the sky.
Nevertheless, they are extremely useful for increasing our knowledge
of the colors of quasars.

\begin{figure}[p]
\epsscale{1.0}
\plotone{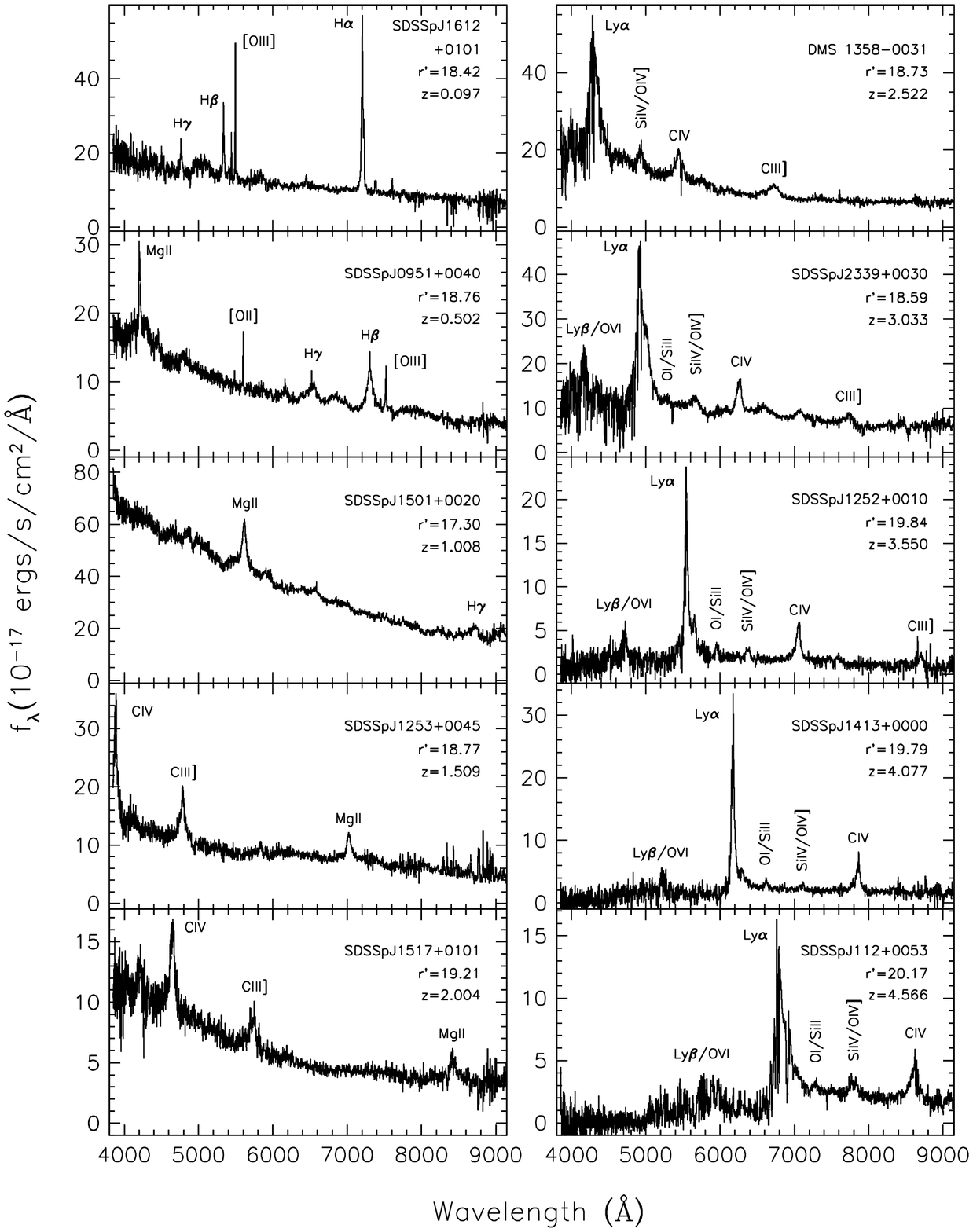}
\caption{Sample SDSS quasars.  These quasars are representative of the
quasars found in the SDSS sample from $z\sim0$ to
$z\sim5$.\label{fig:fig1n}}
\end{figure}

Quasar candidates were observed with the SDSS fiber spectrographs and
the data were reduced with the spectroscopic pipeline \citep{fri+00}.
Each of the quasar identifications and redshifts were determined by
visual inspection by one of us (DEVB), since the automated reduction
pipelines are not yet complete.  The redshifts and photometry of these
1677 objects are not included in Table~\ref{tab:tab2}, but will appear
in 2001 as part of a public SDSS data release.  A sample of 10 SDSS
quasar spectra with $0 < z < 5$ are presented in
Figure~\ref{fig:fig1n}, in order to give the reader an idea of the
quality of the spectra.

\subsection{SDSS Followup}

The remainder of the sample includes 112 quasars discovered during the
course of SDSS followup observations using the ARC 3.5m telescope at
Apache Point Observatory, the Hobby-Eberly Telescope at McDonald
Observatory, and the W. M. Keck Observatory.  These observations are
part of a number of followup projects that are underway, mostly in
search of high-$z$ quasars.  Quasar candidates are selected by color
and confirmed spectroscopically on these other telescopes in order to
confirm objects that the SDSS spectrographs might not observe (e.g.,
due to faintness, fiber spacing constraints, etc.) or that the SDSS
has simply not taken a spectrum of yet.  Those SDSS followup objects
used in this study have been published by \citet{fss+99},
\citet{fss+00}, \citet{shf+00}, \citet{tz00}, \citet{fan+00}, and
\citet{sch+01}.  The redshift and color distribution of this sample of
quasars is summarized in the last of these references.

Nearly all of the $z\ge3.6$ quasars in our sample are from these SDSS
followup observations (which cover a larger area of sky than the SDSS
spectroscopic commissioning plates).  The selection of these quasars
introduces a significant feature in the redshift distribution of our
sample: certain regions ($3 < z < 3.6$) of color space are apparently
underrepresented because of the way that high-redshift follow-up
targets were selected.  The selection criteria for the $z\ge3.6$
quasars are presented by \citet{fan+00} and references therein.

\subsection{Combined Catalog}

The combination of these subcatalogs results in a total of 16,446
quasars covering the entire sky, a fraction of which are in the region
covered by the SDSS commissioning data from the Celestial Equator.  The
distribution of quasars is as follows: 12,987 are from the NASA
Extragalactic Database (NED), 941 are FIRST quasars, and 2518 were
discovered or recovered as part of the SDSS.  The quasars found using
SDSS data are from both the spectroscopic commissioning data (2406
quasars) and from SDSS followup observations using other telescopes
(112 quasars).

In order to create a combined catalog that includes each quasar only
once, the NED, FIRST and SDSS subcatalogs were compared to each other.
Quasars in more than one catalog whose positions agreed to $5\arcsec$
or better, whose redshifts agreed to $\Delta z = 0.1$ or better, and
whose differential magnitude differed by less than 3.5 mag were
considered to be matches.  In all, 508 quasars were found in more than
one catalog of quasars, resulting in a total of 15,938 unique quasars
in the full-sky catalog.

\subsection{SDSS Matches to the Combined Catalog}

In four SDSS Equatorial scans studied, 2625 previously known or
recently discovered quasars from the combined full-sky catalog of
15,938 quasars are found in the SDSS database.  Quasars from the
combined catalog were considered to be matches to the SDSS database if
the positions of the quasars and the SDSS objects agreed to better
than $3\arcsec$ and the agreement between the cataloged magnitudes and
the SDSS magnitudes were reasonable (root-mean-square of the
difference between the predicted and observed $g'$ and $r'$ magnitudes
less than 2 magnitudes, where the predicted SDSS magnitudes have been
computed by transforming the published magnitude to the SDSS system
assuming a spectral index of $\alpha_{\nu}=-0.5$).  The SDSS
astrometry is good to $0.2\arcsec$ or better per coordinate.  See
\citet{fif+00}, Figure 2, for a comparison of the SDSS astrometry that
demonstrates this level of accuracy.  An additional 235 known quasars
had potential matches, but the positions disagreed by more than
$3\arcsec$.  Many of these matches with large positional discrepancies
may be correct, but in order to avoid contamination of the empirical
colors studies herein, we do not include these in our analysis.

\begin{figure}[p]
\epsscale{1.0}
\plotone{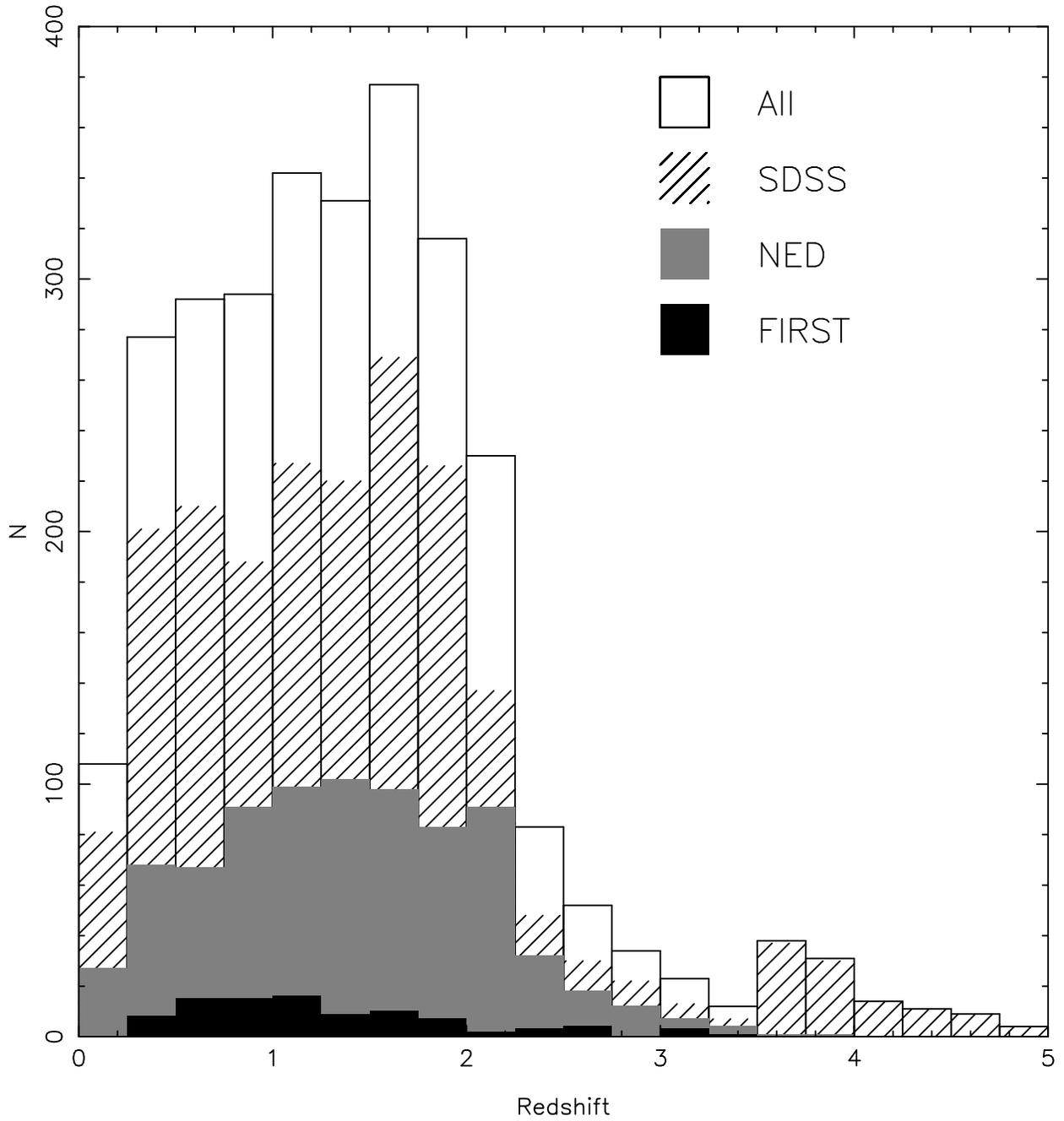}
\caption{Redshift distribution for 801 NED quasars, 92 FIRST quasars,
and 1983 SDSS quasars.  Some quasars appear in more than one sample.
There are 2625 quasars in total.\label{fig:fig2n}}
\end{figure}

Of these 2625 quasars, 801 are from NED (573 unique), 92 are from
FIRST (50 unique), and 1983 quasars are from SDSS data (1759 unique).
A total of 243 quasars were found in more than one of the above
catalogs.  There are more NED quasars in the area covered by these
four runs; however, only these 801 had NED coordinates and photometry
accurate enough to ensure that the matches to SDSS objects are
correct.  A more detailed (and labor intensive!)  analysis would allow
for the matching of additional known quasars and for the correction of
their published coordinates.  A histogram of the redshift distribution
of all 2625 quasars is given in Figure~\ref{fig:fig2n}.  There is a
noticeable lack of quasars between $z=2.5$ and $z=3.6$, since the
density of quasars at $z\ge2.5$ is declining and the followup
observations specifically target quasars at $z\ge3.6$.  The new SDSS
quasars represent over a factor of two increase in the density of
known quasars on the Celestial Equator.

\begin{figure}[p]
\epsscale{1.0}
\plotone{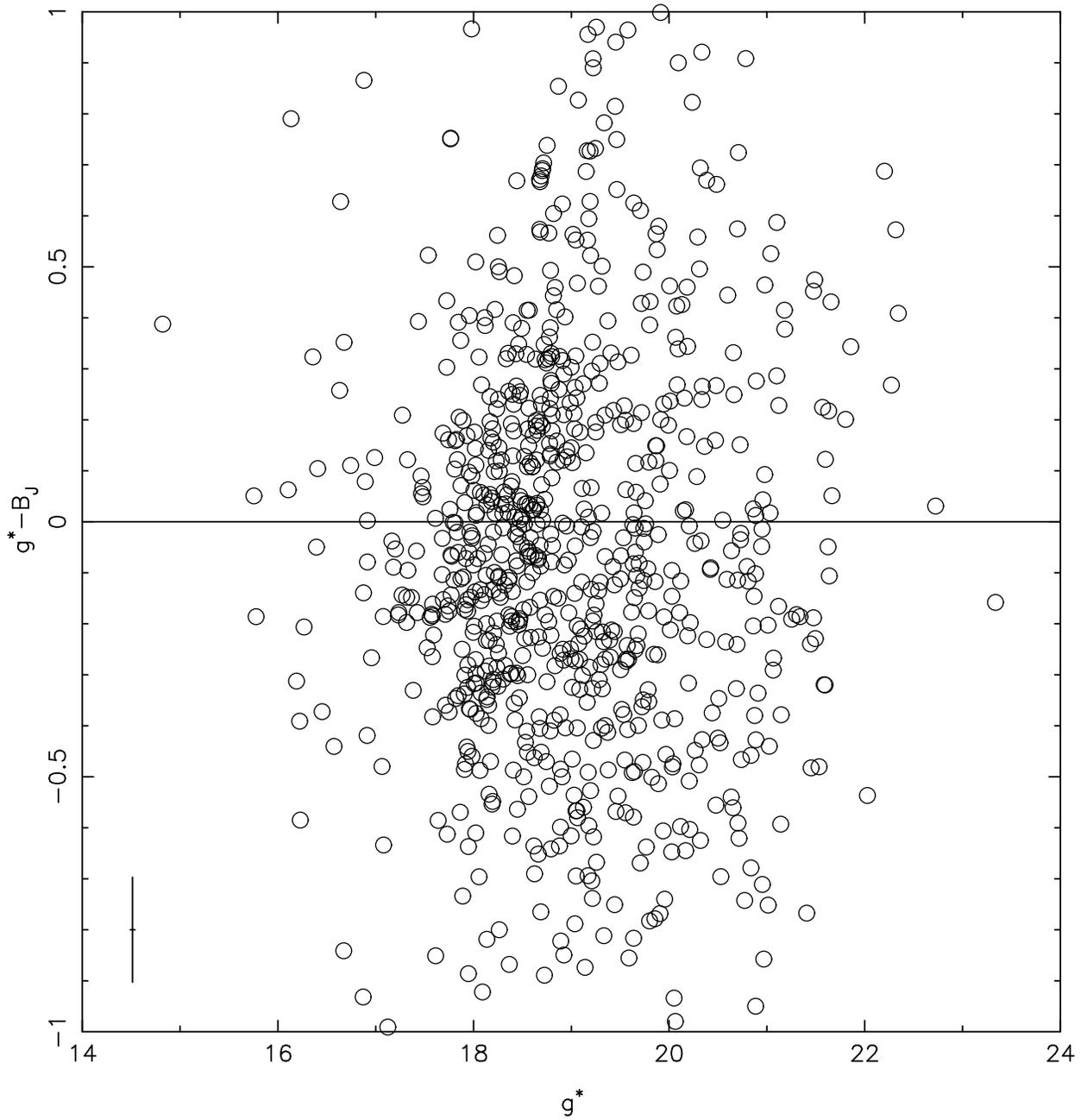}
\caption{Magnitude difference versus magnitude for 867 NED and FIRST
quasars.  For NED quasars the cataloged magnitude is typically, but
not always, $B_J$; for FIRST quasars $B_J$ is replaced by $O$. The
cross in the lower left-hand corner shows a 3\% error in $g^*$ and the
error in $g^*-B_J$ given a 10\% error in $B_J$.
\label{fig:fig3n}}
\end{figure}

The difference between the input magnitude (typically, but not always
$B_J$) and the measured $g^*$ magnitude for the NED and FIRST sources
are plotted as a function of $g^*$ in Figure~\ref{fig:fig3n}.  This
plot enables us to check the accuracy of the cataloged magnitudes and
to look for variability \citep[e.g.,][]{fra96}.  Photometry and
astrometry for all previously published quasars (including SDSS
quasars discovered during followup observations) is presented in
Table~\ref{tab:tab2}.  Data on the 1677 new SDSS quasars will be
published separately as part of an official SDSS data release.  Data
on the 50 FIRST equatorial quasars was kindly provided in advance of
publication by R. Becker and will be presented by \citet{bec+01}.  The
columns in Table~\ref{tab:tab2} are as follows: (1) is the name of the
quasar.  (2) is the redshift.  The right ascension and declination as
measured by the SDSS are given as J2000 coordinates in (3) and (4).
The difference between the SDSS position and the cataloged position is
given in (5) in units of arcseconds.  (6) through (10) give the
measured magnitudes and errors in the five SDSS passbands.  Errors are
photometric errors only and do not include systematic errors, which
are on the order of $0.03$ mag.  Reddening corrections have not been
applied, but the reddening vector in E($g^*$- $r^*$) (as determined
from \citealt{sfd98}) is given in (11).  The last column indicates the
source catalog of each object.  A ``1'' indicates that the object is
in the NED catalog.  SDSS quasars discovered during followup
observations are represented by a ``2''.  Objects found in more than
one catalog are so indicated.

Three quasars were both NED and SDSS sources, and had large
discrepancies in their redshifts as reported by NED.  The following
quasars from NED have incorrect redshifts: UM~203, UM~183, and UM~427.
Their correct redshifts as derived from SDSS spectroscopic
commissioning data are 1.47, 1.14, and 1.69, respectively.  We further
note that PC~0036+0032~($z=4.51$) is not included because it is
improperly classified by NED as a galaxy.  Since not all previously
known quasars in this area of the sky have new SDSS spectra, it is
likely that there are other known quasars in this region with
incorrect redshifts, or that are misclassified.

\section{Quasar Colors} 

\begin{figure}[p]
\epsscale{1.0}
\plotone{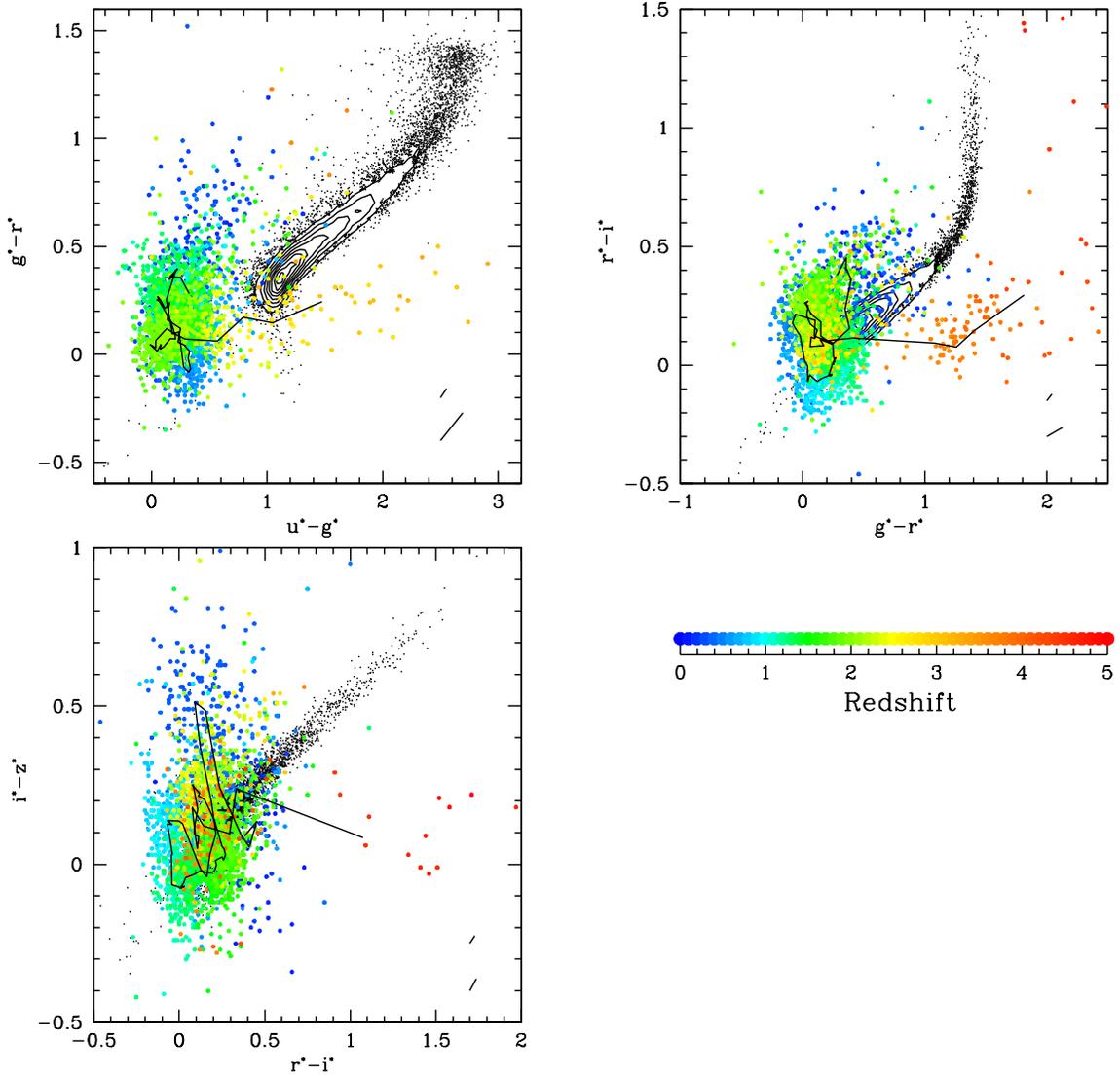}
\caption{SDSS color-color diagrams for 2625 quasars (color points) and
10,000 stars (black points and black contours).  Only objects with
errors less than $0.2$ mag in each band are shown.  Quasar points are
coded as a function of redshift, where the redshift is given by the
color as indicated in the legend.  The long solid black lines are the
median color-color tracks of the quasars.  Galactic (upper) and
extra-Galactic (lower) reddening vectors are given in the lower
right-hand corner of each panel, where the Galactic reddening vector
is the mean over the whole sample.\label{fig:fig4n}}
\end{figure}

We present SDSS quasar color-color diagrams in Figure~\ref{fig:fig4n},
where we plot ($u^*$-$g^*$, $g^*$-$r^*$), ($g^*$-$r^*$, $r^*$-$i^*$),
and ($r^*$-$i^*$, $i^*$-$z^*$), respectively.  The black points and
black countours are 10,000 stellar sources brighter than $i^*=19$
taken from the run 745 data [Galactic ($l,b$) =
(248$^{\circ}$,48$^{\circ}$) to (18$^{\circ}$,27$^{\circ}$)], which
covers approximately the same region of space as run 756.  The color
points are the quasars, color-coded according to their redshifts: blue
points are low-redshift quasars and red points are high-redshift
quasars.  Only quasars whose errors are small ($\le 0.2$ magnitudes
each of the three relevant bands) are plotted.  The solid black line
is the median color-color track of the entire sample.  The average
Galactic reddening and an extra-Galactic (internal) reddening vector
(defined in \S~\ref{sec:red}) are also plotted in the bottom
right-hand corner of each panel in Figure~\ref{fig:fig4n}.  In each
case, the upper-most vector is the average Galactic reddening vector.
Throughout the paper all figures and tables (except
Table~\ref{tab:tab2}) use magnitudes and colors that have been
dereddened using the reddening maps of \citet{sfd98}.
	
\begin{figure}[p]
\epsscale{1.0}
\plotone{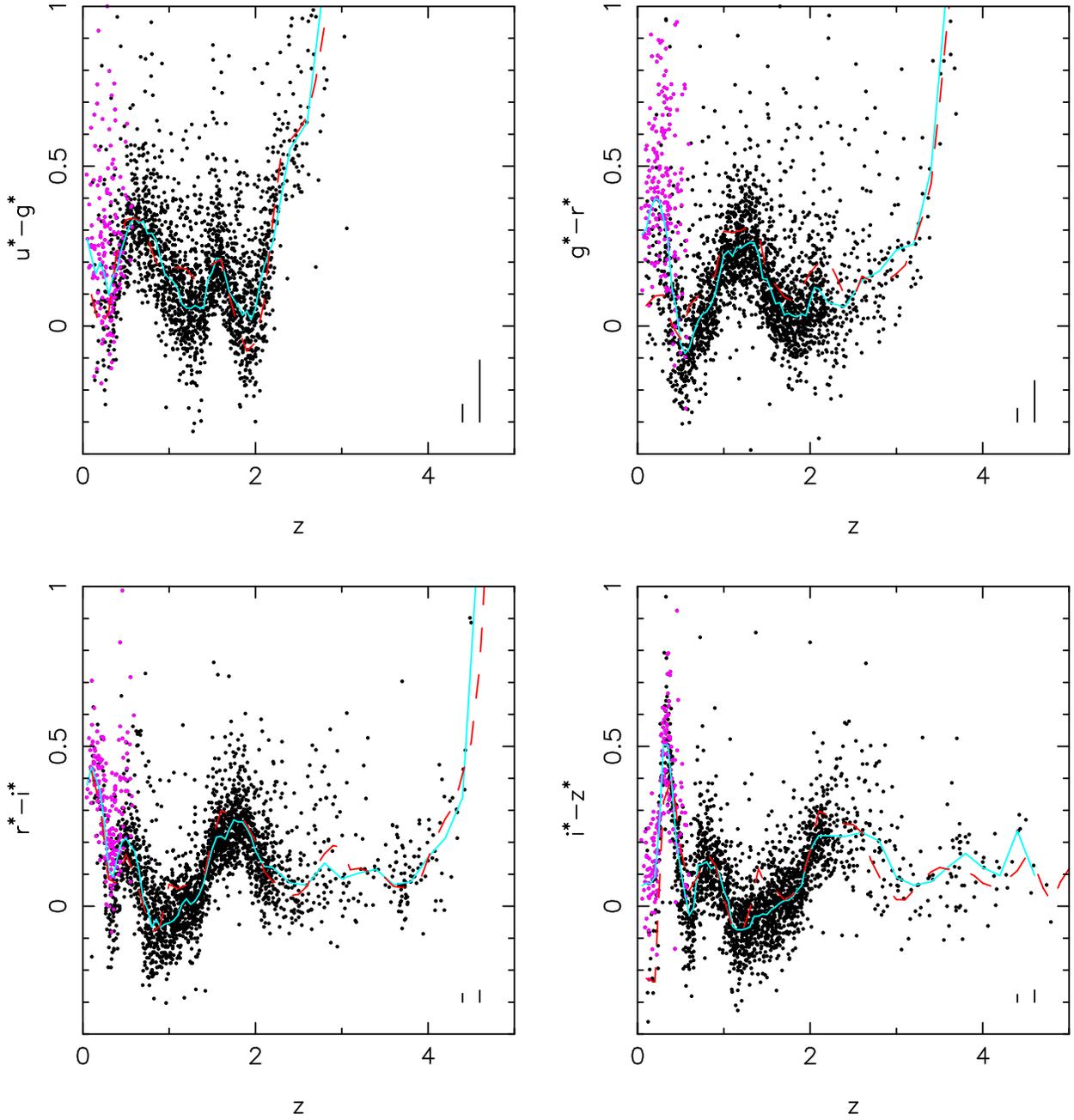}
\caption{The SDSS color versus redshift relation for 2625 quasars
(black dots are point sources, magenta dots are extended sources with
$z\le0.6$).  Only objects with errors less than $0.1$ mag in all bands
are shown.  The median in redshift bins of 0.05 ($z\le2.2$) and 0.2
($z>2.2$) is given by the solid light blue line (smoothed by 50\% of
the bin size).  The dashed red line is a modified version of the
simulated median quasar color-redshift relation from Figure 12 of
\citet{fan99}.  Galactic (leftmost) and extra-Galactic (rightmost)
reddening vectors are given in the lower right-hand corner of each
panel.\label{fig:fig5n}}
\end{figure}

The broad-band SDSS colors of quasars as a function of redshift are
presented in tabular form in Table~\ref{tab:tab3} and in graphical
form in Figure~\ref{fig:fig5n}.  Only quasars whose errors are less
than 0.1 mag ($10\sigma$ detections) in each of the magnitudes that
contribute to a given color are included.  In each redshift bin, we
measure the median colors of the quasars, as well as the limits within
which 95\% of the quasars lie, which we hereafter refer to as
``confidence limits''.  Table~\ref{tab:tab3} gives the median and 95\%
confidence limits of the dereddened colors of quasars in the redshift
bins given, plus the number of quasars per bin.  In
Figure~\ref{fig:fig5n}, the black dots are point sources as seen by
SDSS, whereas magenta dots are extended sources (i.e., the profile is
not suitably described by a point-spread function, as determined by
the SDSS photometric pipeline) with $z\le0.6$.  The solid light blue
line is the median color-redshift vector for these quasars in redshift
bins of size $\Delta z = 0.05$ from $z=0.05$ to $z=2.1$, as reported
in Table~\ref{tab:tab3}.  Starting with $z=2.2$ the bin size is
$\Delta z = 0.2$; the $z=2.2$ bin partially overlaps the $z=2.1$ bin.
The median color-redshift relation in Figure~\ref{fig:fig5n} is
smoothed by 50\% of the bin size for plotting purposes.  We use the
median to describe the color-redshift vector rather than a mean, since
the median is less sensitive to outliers.  The dashed red line is
similar to the median color-redshift track from \citet{fan99}; see
\S~\ref{sec:comp} for further discussion.  Two vectors are plotted as
solid lines in the lower right-hand corner of each panel.  The one on
the left shows the magnitude of typical Galactic reddening, whereas
the one on the right is representative of the change in color due to
extra-Galactic reddening.  These are the same vectors as in
Figure~\ref{fig:fig4n}.  We discuss reddening in further detail in
\S~\ref{sec:red}.

The high quality of the data is quite apparent in
Figure~\ref{fig:fig4n} and can be compared to similar graphs using
photometry from photographic plates \citep[e.g.,][]{who91,imh91}.  The
tightness of the stellar locus is a result of the high-quality CCD
photometry.  Clearly, more accurate photometric data allow quasars to
be found more easily.  A lack of color-degeneracy in the quasar colors
as a function of redshift is apparent in the color-color
(Figure~\ref{fig:fig4n}) and the color-redshift diagrams
(Figure~\ref{fig:fig5n}).  Quasars with similar redshifts tend to have
the same colors, whereas quasars with different redshifts occupy
different places in color space.  The fact that there is so much
structure in the color-redshift relation and that the scatter in the
colors at a given redshift is reasonably small may allow for the
determination of photometric redshifts for quasars \citep{rfs+00}.
Photometric redshifts are now common practice for galaxies
\citep[e.g.,][]{bsc+00}, and have recently become possible for quasars
\citep{wmr+00}.

\subsection{Theoretical Colors}

We begin our analysis with a discussion of the expected colors of
quasars in the SDSS photometric system.  For a general introduction on
the subject see \citet{fc94} and \citet{who94}, or \citet{fan99} for a
discussion specific to the SDSS.  To first order, the spectra of
quasars can be characterized as a power-law.  A power-law spectrum has
the convenient property that it has the same color at any redshift.
The color of an object with flux density $f(\nu)$ is given by:
\begin{equation}
m_1 - m_2 = -2.5\left(\log\frac{\int f(\nu)S_1(\nu)d\log\nu}{\int S_1(\nu)d\log\nu} - \log\frac{\int f(\nu)S_2(\nu)d\log\nu}{\int S_2(\nu)d\log\nu}\right),
\end{equation}
where $S_1(\nu)$ and $S_2(\nu)$ are the throughput of the system in
each bandpass (for a photon counting system).  For the SDSS
photometric system, there is no additive constant since the SDSS
magnitudes are on the $AB$ system \citep{og83,fig+96}.  This lack of
an additive constant is in contrast with the $UBVRI$ system, which is
Vega based; the computation of colors in the $UBVRI$ system requires a
correction for the fact that the spectrum of Vega is not perfectly
flat in $f_{\nu}$.  For a flat-spectrum source ($\alpha_{\nu} = 0$),
the colors in an $AB$ system are always $0$.  Table~\ref{tab:tab4}
gives the expected SDSS colors for a range of spectral indices.  The
spectral indices are given in both frequency and wavelength units,
where $\alpha_{\lambda}$ and $\alpha_{\nu}$ are defined such that
$\alpha_{\lambda} = -(2+\alpha_{\nu})$, $f_{\nu} \propto
\nu^{\alpha_{\nu}}$ and $f_{\lambda} \propto
\lambda^{\alpha_{\lambda}}$.  Expected colors for other spectral
indices can be interpolated or extrapolated from those given in
Table~\ref{tab:tab4}.

From Figure~\ref{fig:fig5n} it is clear that the colors of quasars are
not strictly power-laws; however, the average colors are consistent
with the input power-law distribution of \citet{fan99}, $\alpha_{\nu}
= -0.5\pm0.3$.  Whereas the power-law index of the quasar spectrum
sets the average colors of a quasar, the colors will deviate from this
value as a result of emission and absorption features (e.g. the
Lyman-$\alpha$ forest).  To identify the nature of these
discrepancies, we turn to an analysis of the expected colors of
quasars using a composite SDSS quasar spectrum convolved with the SDSS
transmission curves.

\subsection{Composite Spectrum Colors}

A composite quasar spectrum has been constructed from the early SDSS
spectroscopic commissioning spectra \citep{vbr00,van+01}.  The
composite spectrum extends from $1050\,{\rm \AA}$ to $7000\,{\rm
\AA}$, in the rest frame.  We have convolved this spectrum with the
most recent SDSS transmission curves \citep[Appendix
A]{fan+00}.  The resulting color-redshift tracks for the composite
spectrum are presented in Figure~\ref{fig:fig6n}.

In Figure~\ref{fig:fig6n} the solid blue line is the color-redshift
track of the composite spectrum.  For comparison, we also re-plot the
empirical median color-redshift tracks from Figure~\ref{fig:fig5n}
(solid black line) and the expected colors from a power-law spectrum
of the form $f_{\nu} \propto \nu^{-0.5}$ (dashed black line; see
Table~\ref{tab:tab4}).  The remaining lines in Figure~\ref{fig:fig6n}
show what happens when a given emission feature is cut out of the
composite spectrum and is replaced with a power-law, i.e. each
emission line is removed from the composite spectrum, then the flux
values in the wavelength range removed are replaced by a power-law
spectrum of the form $f_{\nu} \propto \nu^{-0.3}$ normalized at
$1450\,{\rm \AA}$, which is a good fit to the composite spectrum
\citep{vbr00,van+01}.  A small additive constant was also required to
make the spliced region fit into the spectrum cleanly.  Since the
strength of the Lyman-$\alpha$ forest is a function of redshift and
because the composite does not cover the entire optical spectrum for
$0 < z < 5$, the ends of the composite spectrum are padded with the
average value between $1050$ and $1150\,{\rm \AA}$ at the blue end,
and $7000$ and $8000\,{\rm \AA}$ at the red end.  As a result, the
absolute colors may not be accurate in the blue colors at high
redshift and in the red colors at low redshift.
Figure~\ref{fig:fig6n} displays what happens when Lyman-$\alpha$,
\ion{C}{4}, \ion{Mg}{2}, H$\beta$, H$\alpha$, or the small blue
($\lambda\,3000$) bump (SBB) are removed.  In \S~\ref{sec:comp}, we
will demonstrate the effect that a strong emission line has on the
colors of an otherwise power-law spectrum.  A knowledge of this effect
is helpful to understanding the results in the next section.

\subsection{Empirical Colors}

The color-redshift relations of quasars in the SDSS bandpasses exhibit
considerable structure.  We attempt to describe the causes of the
features in the color-redshift diagrams.  Features in the
color-redshift relations are identified by comparing the results from
the composite spectrum above (Figure~\ref{fig:fig6n}) to the measured
colors of quasars (Figure~\ref{fig:fig5n}).  All colors used for this
analysis are dereddened colors.  The emission lines that significantly
affect the broad-band colors include Lyman-$\alpha$, \ion{C}{4},
\ion{Mg}{2}, H$\alpha$, H$\beta$, and the $\lambda\,3000$ bump.  The
$\lambda\,3000$ bump consists of Balmer continuum emission and
\ion{Fe}{2} emission features \citep{gra82,pet97}.  The feature starts
at approximately $2300\,{\rm \AA}$ and extends to approximately
$3800\,{\rm \AA}$.  The \ion{Mg}{2} emission line is nearly centered
between these ranges, and contributes significantly to color changes as
a function of redshift since it sits on top of the $\lambda\,3000$
bump.

We identify each of the significant features in the color-redshift
curves.  Many of the features in the color-redshift relation are
caused by more than one feature in the quasar spectrum interacting
with the transmission curves.  A more detailed understanding of
the causes of deviations from power-law colors can be obtained by
comparing Figure~\ref{fig:fig6n} to Figure~\ref{fig:fig5n}.
Understanding the causes of the color-redshift features is interesting
in and of itself, but also is helpful for determining sample
completeness for objects with particularly weak or strong emission
features.

\subsubsection{$u'$-$g'$}

$z\sim0.1$ to $0.4$ --- The colors in this region of redshift have a
considerable range; the range is much larger than for the redder SDSS
colors.  A comparison of the composite spectrum colors to the
composite spectrum colors without the $\lambda\,3000$ bump reveals
that some of this range could be due to objects having a wide range of
$\lambda\,3000$ bump strengths, but see \S~\ref{sec:red} for another
explanation.

$z\sim0.3$ --- The $\lambda\,3000$ bump and \ion{Mg}{2} are in $u'$, and
cause the color to be blue.

$z\sim0.6$ --- The $\lambda\,3000$ bump and \ion{Mg}{2} are in $g'$, making
the color red.

$z\sim1.3$ --- \ion{C}{4} is in $u'$ and shifts the color to the blue.

$z\sim1.6$ --- Lyman-$\alpha$ is entering $u'$, while \ion{C}{4} is
leaving $u'$ and entering $g'$.

$z\sim1.9$ --- Lyman-$\alpha$ is in $u'$, but the effect is tempered by
the presence of \ion{C}{4} in $g'$.  The strength of Lyman-$\alpha$
dominates nevertheless, causing a blue dip.

$z\sim2.3$ to $2.4$ --- A small plateau-like feature is caused as
\ion{C}{4} leaves $g'$ shortly after Lyman-$\alpha$ enters $g'$.

$z>2.6$ --- The $u'$-$g'$ color rises rapidly as the Lyman-$\alpha$
forest and Lyman-limit systems cause there to be little or no flux in
the $u'$ band.

\subsubsection{$g'$-$r'$}

$z\sim0.2$ to $0.3$ --- The presence of H$\beta$ in $r'$ keeps the
average color relatively red.

$z\sim0.5$ --- The $\lambda\,3000$ bump is mostly in $g'$, causing the color
to be blue.  As with the $z<0.4$ region in $u'$-$g'$, there is
evidence for a population of redder objects.

$z\sim1.2$ --- The combination of \ion{Mg}{2} and the $\lambda\,3000$
bump in $r'$ cause a reddening near this redshift.

$z\sim1.75$ --- The $\lambda\,3000$ bump and \ion{C}{4} offset each other as
the latter enters $g'$ and the former leaves $r'$.

$z\sim2.1$ --- The presence of \ion{C}{4} in $g'$ keeps the color bluer
than a power-law.

$z\sim2.5$ to $3.5$ --- The \ion{C}{4} line pushes the color to the
red, whereas Lyman-$\alpha$ pushes the color to the blue.  As the
redshift increases more and more of the Lyman-$\alpha$ forest enters
$g'$ and the color reddens.

$z>4.0$ --- The $g'$-$r'$ color rises rapidly as the Lyman-$\alpha$
forest and Lyman-limit systems cause there to be little or no flux in
the $g'$ band.

\subsubsection{$r'$-$i'$}

$z\sim0.1$ to $0.2$ --- H$\alpha$ is in $i'$ and makes the color
redder than the average power-law value.

$z\sim0.3$ --- The presence of H$\beta$ in $r'$ drives the color
blueward, which is enhanced when H$\alpha$ leaves $r'$.

$z\sim0.5$ --- The color moves back to the red while H$\beta$ is in
$i'$.

$z\sim0.9$ --- The $\lambda\,3000$ bump fills the $r'$ filter making the
color bluer.

$z\sim1.2$ --- A small kink in the median colors is noticeable here as
the $\lambda\,3000$ bump and \ion{Mg}{2} trade off in influence in the
$r'$ filter.

$z\sim1.4$ to $1.5$ --- The color makes a sharp transition from blue to
red as \ion{Mg}{2} leaves $r'$ and enters $i'$.

$z\sim1.65$ --- A small kink in the median colors is noticeable here
as the $\lambda\,3000$ bump and \ion{Mg}{2} trade off in influence in
the $i'$ filter.

$z\sim1.8$ --- A small hump is caused as both \ion{Mg}{2} and the
$\lambda\,3000$ bump push the color redward.

$z\sim2.0$ to $2.5$ --- The color is driven back to the blue as the
$\lambda\,3000$ bump leaves $i'$.

$z\sim2.6$ to $3.4$ --- \ion{C}{4} keeps the color blue, when it would
otherwise have reddened.

$z\sim3.4$ to $4.4$ --- Lyman-$\alpha$ and \ion{C}{4} offset each other
during the period when the former is in $r'$ and the latter is in
$i'$.

$z>4.5$ --- The $r'$-$i'$ color rises rapidly as the Lyman-$\alpha$
forest and Lyman-limit systems cause there to be little or no flux in
the $r'$ band.

\subsubsection{$i'$-$z'$}

$z\sim0.2$ --- H$\alpha$ is in $i'$ and keeps the color blue.

$z\sim0.3$ to $0.4$ --- H$\alpha$ has moved into $z'$, causing an
abrupt reddening.

$z\sim0.6$ --- H$\beta$ is in $i'$, whereas H$\alpha$ is leaving
$z'$, resulting in a blue color.

$z\sim0.8$ --- H$\beta$ has moved into $z'$ and causes the color to
become redder.

$z\sim1.2$ --- The $\lambda\,3000$ bump in $i'$ drives the color blueward.

$z\sim1.9$ to $2.1$ --- \ion{Mg}{2} moves from $i'$ to $z'$, causing an
abrupt reddening.

$z\sim2.1$ to $2.5$ --- As with the $z\sim1.7$ feature in $r'$-$i'$, a
small red hump is formed as both \ion{Mg}{2} and the $\lambda\,3000$
bump push the color to the red.

$z\sim3.4$ to $4.2$ --- \ion{C}{4} keeps the color bluer than the average
power-law color.

$z\sim4.4$ to $5.0$ --- Lyman-$\alpha$ and \ion{C}{4} offset each other
as \ion{C}{4} pushes the color redward at the same time that
Lyman-$\alpha$ pushes the color blueward.

\subsection{Comparison of Expected and Measured Colors\label{sec:comp}}

To characterize the differences between the expected colors of a
power-law and the empirical colors of quasars, it is instructive to
know what effect the profile of a strong emission line has on the
colors of quasars as the line is redshifted through each of the
filters.  Figure~\ref{fig:fig7n} shows what happens when a top hat
emission line with an observed equivalent width of $200\,{\rm \AA}$
that is $20\,{\rm \AA}$ wide is redshifted through each of the
filters.  The emission line starts centered at $3000\,{\rm \AA}$
($z=0$) where the $u'$ filter begins and ends at $11,500\,{\rm \AA}$
($z=2.833$) where the $z'$ filter ends.  The four curves are
$u'$-$g'$, $g'$-$r'$, $r'$-$i'$, and $i'$-$z'$, respectively, with
$u'$-$g'$ being the curve in the lower left.  These features are
similar to the effect that the Lyman-$\alpha$ emission line has on the
broad-band colors of quasars.  Note the similarity of the dip in
$u^*$-$g^*$ near $z=1.9$ in Figure~\ref{fig:fig5n} to the dip near
$z=0.2$ in Figure~\ref{fig:fig7n}.

A broad feature, such as is characterized by \ion{Fe}{2} emission, has
a similar profile, although the slope of the color change between
bandpasses is less striking.  It is particularly interesting that a
relatively weak, but broad emission features such as the \ion{Fe}{2}
complexes can have an equal effect to that of strong, narrow emission
feature such as Lyman-$\alpha$.  The total equivalent width within the
bandpass matters much more than the manner in which that equivalent
width is distributed, since the bandpasses are broad.  For example,
note the similarity of the $r^*$-$i^*$ colors from $z=0.6$ to $z=2.4$
and the $i^*$-$z^*$ colors from $z=0.8$ to $z=3.0$ in
Figure~\ref{fig:fig5n} to the $r'-i'$ and $i'-z'$ shapes,
respectively, in Figure~\ref{fig:fig7n}.  These features are produced
by a combination of the $\lambda\,3000$ bump and \ion{Mg}{2} emission.

One of the most powerful uses of this analysis is the refinement of
the simulated quasar spectrum, which, in turn, tells us about the
empirical properties (spectral indices, emission line equivalent
widths, etc.) of quasars.  Fan (1999) calculated theoretical colors of
quasars in the SDSS filter system as a function of redshift.  Although
these simulated colors are correct to first order (they correctly
reproduce the colors of quasars on the ensemble average), there are
deviations from the average that have a significant impact upon the
colors of quasars as a function of redshift.  The simulated quasar
color-redshift track shown in Figure~\ref{fig:fig5n} is similar to
that of \citet{fan99}, but uses a larger equivalent width for the
\ion{Fe}{2} features, and the updated filter curves.  This theoretical
color-redshift relation matches the observed color-redshift relation
surprisingly well.  While the simulation does indeed fit the data
quite well, there are still regions of color-redshift space where the
simulations and the data do not agree well.

\subsubsection{The $\lambda\,3000$ Bump}

One piece of information that we can glean from a comparison of the
simulated color-redshift tracks to the observed tracks is the
structure of the $\lambda\,3000$ bump.  The quality of the digital
photometry provided by the SDSS allows us to determine the structure
of this spectral feature without having to examine any spectra.  Such
an analysis is of particular interest given the role that the
$\lambda\,3000$ bump plays in the spectra of quasars \citep{nww+85}.

Some of the most notable deviations of the empirical colors from the
simulated colors occur at $z=1.0$ and $z=1.4$ in $r^*$-$i^*$ and
$i^*$-$z^*$, respectively.  The sharp break in the simulated colors
from Figure~\ref{fig:fig5n} is caused by the fact that the \ion{Fe}{2}
emission surrounding the \ion{Mg}{2} emission line has been modeled as
two distinct features following \citet{fhf+91}.  However, a comparison
with the empirical color-redshift track shows that the two bumps on
either side of \ion{Mg}{2} actually combine to form an apparent
continuum, such that \ion{Mg}{2} lies well above the power-law
continuum level.  As a result, the regions surrounding \ion{Mg}{2}
emission are not representative of the power-law continuum spectrum.
In addition to the merging of what \citet{fhf+91} call
``\ion{Fe}{2}~feature~2'' and ``\ion{Fe}{2}~Feature~3'', it is also
quite possible that ``\ion{Fe}{2}~Feature 1'' (longward of \ion{C}{4}
emission) merges with both of these, and that the dip near $2200\,{\rm
\AA}$ in the flux of composite quasar spectra is caused by dust
(\citealt{rs80}, but see \citealt{osk84} for a counter-argument).

The effect of the $\lambda\,3000$ bump at $z\le0.6$ is also quite
evident.  For $r^*$-$i^*$ and $i^*$-$z^*$ the colors of quasars are
very tight at $z\le0.6$.  There is considerable variation as a
function of redshift, but very little scatter at a given redshift.
This is not the case for $u^*$-$g^*$ and $g^*$-$r^*$, where there is a
large range of colors.  A possible explanation is that quasars can
exhibit a broad range of strength in the $\lambda\,3000$ bump at low
redshifts.  A closer examination of the definition of a quasar at low
redshift is warranted (see \S~\ref{sec:seyferts} for more details and
another possible explanation).

\subsubsection{\ion{C}{4}}

There are significant differences between the simulated and measured
colors of quasars in the redshift ranges affected by \ion{C}{4}.  Note
in particular the $z\sim1.3$ region in the $u^*$-$g^*$ panel and the
$z\sim2.1$ region in the $g^*$-$r^*$ panel of Figure~\ref{fig:fig5n}
and Figure~\ref{fig:fig6n}.  A possible explanation for these
discrepancies is that the assumed equivalent width for \ion{C}{4} is
considerably smaller than the average equivalent width of \ion{C}{4}
in our sample.  This is surprising because the \ion{C}{4} equivalent
width in the simulation is taken from the LBQS composite quasar
spectrum \citep{fhf+91}, and LBQS quasars contribute significantly to
our sample.  However, as with \ion{Mg}{2} and the $\lambda\,3000$
bump, \ion{C}{4} may be influenced by other emission, in particular
lines of \ion{Fe}{2} ($\lambda 1700$ to $2200\,{\rm \AA}$),
\ion{He}{2} $\lambda 1640$, and \ion{O}{3}] $\lambda 1663$.  To the
extent that these lines are missing from the simulations, the
simulations will deviate from the empirical colors.

\subsubsection{H$\alpha$}

A particularly useful diagnostic can be made from the effect of
H$\alpha$ on the broad-band colors of SDSS quasars.  That this is the
case can be seen at $z\sim0.2$ in the $i^*$-$z^*$ of
Figure~\ref{fig:fig5n}.  The deviation of the simulated colors of
\citet{fan99} from the observed colors is caused by the fact that the
simulations used the original transmission curves.  It is now known
that the true transmission curves deviate somewhat from those
originally reported \citep{fan+00}.  The sharpness of the
color-redshift feature in $i^*$-$z^*$ due to H$\alpha$ can be used as
a diagnostic throughout the course of the Survey.  Small changes in
the transmission curves can be monitored as a function of time and as
a function of CCD chip in the camera.

\section{Discussion}

The primary purpose of this investigation of the empirical colors of
quasars in the SDSS photometric system is to aid in the selection of
quasars for spectroscopic observations during the course of the Sloan
Digital Sky Survey.  There are a number of complex issues involved in
the selection of quasars in the most complete and efficient manner.
The data presented herein provides a wealth of information that can be
used to address a number of topics.  We now turn to a discussion of
some of the more important issues.

\subsection{Reddening\label{sec:red}}

Although the photometry presented herein has been corrected for
Galactic reddening using the reddening maps of \citet{sfd98}, it is
important to understand how much effect Galactic reddening has on the
colors of quasars.  The average Galactic reddening vector for the
quasars in our sample is ($0.055, 0.042, 0.027, 0.024$) in
($u^*$-$g^*$, $g^*$-$r^*$, $r^*$-$i^*$, $i^*$-$z^*$) coordinates.
These vectors are plotted to scale as the left-most of the two vectors
in the lower right-hand corner of the color-redshift curves in
Figure~\ref{fig:fig5n} and the upper-most of the two vectors in the
lower-right hand corner of the color-color plots in
Figure~\ref{fig:fig4n}.  At low Galactic latitudes, Galactic reddening
can be considerably more significant (c.f., \citealt{fan99}, Figure
5).

In contrast to Galactic reddening, little is known about inter-galactic
reddening and reddening internal to quasars.  Although it has been
suggested that a significant fraction of quasars may be reddened, and
therefore missed in flux-limited surveys \citep[e.g.,][]{fp93,fww99},
it is difficult to know exactly how prevalent internal reddening is
and to what extent it effects the broad-band colors of quasars.  Close
examination of the color-redshift relations in Figure~\ref{fig:fig5n}
reveals that the spread in colors around the median color as a
function of redshift is much larger for $u^*$-$g^*$ than for the other
colors.  Moreover, the scatter is asymmetric, in the sense expected
for reddening.

To quantify this effect, we subtract the median colors given in
Table~\ref{tab:tab3} at each redshift from the colors of the quasars
shown in Figure~\ref{fig:fig5n} (i.e., those with small photometric
errors) with $0.4 < z < 3.0$, and examine the residual colors.
Quasars that are classified as extended (as opposed to point sources)
are excluded to avoid contamination from the host galaxy.
Low-redshift objects are excluded to avoid confusion with Seyfert
galaxies.  High-redshift objects are excluded due to a relative
paucity of data.  Histograms of the residual colors are given in
Figure~\ref{fig:fig8n}.  Note the red tail in the distribution of
$u^*$-$g^*$, and, to a lesser extent, $g^*$-$r^*$.  In order to
determine the cause of this tail, we have created additional
diagnostic plots.

The residual colors are plotted as a function of magnitude in the four
panels of Figure~\ref{fig:fig9n}, as a function of redshift in the
four panels of Figure~\ref{fig:fig10n}, and as a function of absolute
magnitude ($q_{\rm o}=0.5$, $H_{\rm o}=65\,{\rm km}\,{\rm
sec}^{-1}\,{\rm Mpc}^{-1}$) in Figure~\ref{fig:fig11n}.  Except for
Figure~\ref{fig:fig10n} and Figure~\ref{fig:fig11n},
Figures~\ref{fig:fig8n} to \ref{fig:fig12n} exclude all quasars with
$z<0.4$.  Two-sigma error bars for objects with magnitude 20 (in the
bandpass given) are plotted in Figure~\ref{fig:fig9n}.  Note that the
scatter blueward of the median colors ($\equiv 0$) for each pair of
filters is less than the amount of scatter redward of the median
colors.  This scatter increases towards the red with fainter
magnitudes, is more pronounced in the bluer colors, and is independent
of redshift.

Such an effect is quite likely the result of reddening of the quasars,
either external or internal, but certainly extra-Galactic; the fact
that the reddening is independent of redshift points towards internal
reddening over external reddening, since external reddening would
depend on the volume of space enclosed at a given redshift.
Table~\ref{tab:tab5} gives the blue and the red one-sided 95\%
confidence deviations in each of the median-corrected colors for those
objects brighter than $i^*=21$ and whose errors in each band are less
than $0.1$ mag.  Also given are the corresponding values of spectral
index.

We further demonstrate the reddening of some objects in
Figure~\ref{fig:fig12n} where we plot the corrected colors against
each other; typical errors are given in the upper right hand corner of
each panel.  The vector in each of the panels is the absolute value of
the {\em blue} 95\% confidence limit from Table~\ref{tab:tab5}.
Objects whose red color excess is comparable to or larger than the
absolute value of the blue 95\% confidence limit given in
Table~\ref{tab:tab5} show signs of reddening beyond what is expected
from Galactic dust.  The difference between the absolute value of the
blue 95\% confidence limit given in Table~\ref{tab:tab5} and the red
95\% confidence limit given in Table~\ref{tab:tab5} yields an estimate
of the amount of extra-Galactic reddening.  The resulting vector,
(0.193, 0.129, 0.037, 0.038), is the extra-Galactic reddening vector
used in Figure~\ref{fig:fig4n} and Figure~\ref{fig:fig5n}.  This
vector is similar in direction to the Galactic reddening vector.

We have defined a sample of 26 ``reddened'' quasars, which are those
quasars whose red color excess in $u^*$-$g^*$, $g^*$-$r^*$, and
$r^*$-$i^*$ lie outside the blue 95\% confidence interval.  Nearly
half of these are radio sources, as we discuss below in
Section~\ref{sec:radio}.  Some examples of these ``red'' quasars are
given in Figure~\ref{fig:fig13n}, which presents 10 of the 20 red
quasars for which we have SDSS spectra.  We show these spectra in
order to demonstrate that these objects are truly redder than the
average and are not simply the result of photometric errors, etc.
Note that these spectra contain some Broad Absorption Line (BAL)
quasars and many have strong, narrow absorption lines.  These spectra
can be compared to those presented in Figure~\ref{fig:fig1n}, which
shows the spectra of 10 more normal SDSS quasars spanning a range in
redshift from $z\sim0$ to $z\sim5$.

The extent of the reddening in these quasars is a function of the
assumed reddening curve.  Whether the reddening is internal or
external, features such as the $2200\,{\rm \AA}$ bump \citep{mat94}
which is typically associated with reddening from dust grains, may
influence the colors in a non-linear manner.  If the reddening is
internal, it may come from depletion by material associated with the
torus of gas and dust that is thought to surround the central quasar
engine; if this is the case, reddening might be used as an orientation
indicator for quasars that are not radio-detected.  Reddening might
also arise in the quasar host galaxy or in other galaxies along the
line of sight.

Reddened quasars can have colors that can be well removed from the
predicted location of quasars in color-color space.  Color-selected
surveys may select against such quasars.  Note in particular the
dearth of faint red quasars in the upper right-hand panel of
Figure~\ref{fig:fig9n}.  If there is a significant population of
reddened quasars, they may constitute the optical counterparts of the
remainder of the hard X-ray background that have hitherto gone
undetected \citep{bhs+00,mcb+00}.

\subsection{Spectral Index Distribution}

Table~\ref{tab:tab5} gives the blue and red one-sided range in
spectral index (95\% confidence) needed to produce the observed range
of colors.  If the blueward scatter in the colors is representative of
the true (unreddened) scatter in optical spectral index, the spectral
index distribution of the sample is approximately
$\alpha_{\nu}=\pm0.65$ (95\%~confidence), where the error gives the
range of values that includes 95\% of sources with colors bluer than
zero.  Note that we have not determined the average spectral index
from the data, but instead use $\alpha_{\nu}=-0.5$, which is typically
used for the optical spectral indices of quasars and is not a bad fit
to the data \citep{van+01}.  A determination of the exact value of the
average spectral index would require a large region of quasar spectra
to be devoid of emission lines, which is not the case in the
UV/optical part of the spectrum.

Another conclusion that may be drawn from Figure~\ref{fig:fig10n} is
that the scatter in the colors of quasars at a given redshift must be
primarily due to the range of optical spectral indices of the
individual quasars.  If the scatter in the colors was instead
dominated by the strength of emission lines, then we would expect that
this scatter would be a stronger function of redshift.  In addition,
as noted by \citet{who94}, a large intrinsic range of quasar spectral
indices will blur the calculations of the luminosity function.
Extrapolation of a magnitude to a different band using the wrong
spectral index will give the wrong magnitude in the new band.  For the
observed range of spectral indices, the range of absolute $g^{*}$
magnitudes at $z=2$ given a selection function based on observed
$i^{*}$ magnitudes is $\Delta M_{g'} = 0.89$.

\subsection{Radio Sources\label{sec:radio}}

Although the vast majority of the quasars that the SDSS finds will be
color-selected quasars, many will be radio-selected.  All FIRST
sources with point-like optical counterparts and $i^*<19$ will be
targeted as quasar candidates.  Targeting these radio-detected sources
will assist in the determination of the completeness of our
color-selected sample; using these objects, we will be able to test
whether a significant fraction of quasars is being missed.  At the
present time, however, our sample of radio-detected quasars is not
large enough to determine if the color-redshift distribution of these
quasars are significantly different from that of color-selected
objects.

We can, however, comment on the ensemble average of the colors of
these quasars.  For radio sources only, the blue scatter in the median
corrected colors is consistent with scatter of $\pm0.64$ in the
spectral index (95\% confidence).  The scatter of the colors to the
blue is not significantly different from the distribution of the
sample as a whole.  On the other hand, the 95\% confidence limit in
the red scatter requires spectral indices redder than $\alpha_{\nu} =
-1.5$, which is considerably redder than the sample as a whole.  This
result is consistent with the fact that many (12 of 26) of the
reddened quasars discussed above are FIRST radio sources.

\subsection{UVX Color Selection}

Since the colors of $z<2.2$ quasars are dominated by the continuum
power-law spectrum, it is expected that the UVX selection technique
for these quasars should be relatively complete, i.e. their median
colors are generally significantly bluer than the $u'$-$g'$ selection
criterion.  However, from Figure~\ref{fig:fig5n} we see that the
$u^*$-$g^*$ color of $z\le2.2$ quasars can, in fact, be as red as or
redder than $u^*$-$g^*\sim0.6$.  Quasars with $z\sim0.1$, $0.6$,
$1.6$, and $2.2$ can have colors close to the UVX cutoff.  As such, it
is {\em not} possible to assume that the UVX technique is complete.
Flux-limited samples will indeed be incomplete near the flux limit as
a result of larger scatter in the colors at the faint end of the
distribution, and this incompleteness is a function of redshift.  For
example, \citet{hv93} have suggested that the fall-off in the number
density of faint quasars as found by UVX quasar surveys such as
\citet{bsp88} could be a result of this effect.  One way to test this
is to conduct a survey of faint quasar candidates to a limiting $g'$
magnitude of $\sim 22$ (comparable to previous surveys) and compare
the luminosity function of quasars in redshift regimes where median
$u'$-$g'$ colors are much bluer than the UVX cutoff to those quasars
in redshift regimes where median $u'$-$g'$ colors are very close to
the UVX cutoff.  Such a project would have to be conducted as a
follow-up study to the SDSS using the data from the SDSS equatorial
region in the South Galactic Cap which will be imaged multiple times.

Additional problems with UVX quasar selection arise as a result of the
colors of stars.  At very bright magnitudes ($g^*\sim 16$) there is
very little stellar contamination in the UVX regime.  However, at
fainter limits, there is significant contamination of UVX quasars by
stars, due to the metal-poor halo population.  Figure~\ref{fig:fig14n}
plots a color-magnitude diagram of blue stellar sources (as black
contours and black points) from run 756, camera column 3.  At the
bright end, the majority of the blue objects will be quasars, with
some contamination from white dwarfs.  As $g^*$ gets fainter, there
are more and more stars that have blue colors.  This is partly due to
increasing photometric error at fainter magnitudes, but it is also
likely to be the result of shifts in the stellar locus as a function
of magnitude (and therefore metallicity) as discussed by
\citet{nrr+99} and \citet{fif+00}.  The blue extension of the contours
would only be half as large if the observed effect was purely due to
larger photometric errors at fainter magnitudes.  The horizontal line
in Figure~\ref{fig:fig14n} gives the $g^*$ magnitude for a limiting
magnitude of $i^*=19$ assuming a power-law spectrum of $\alpha_{\nu} =
-0.5$.  The vertical line gives an approximate cutoff for UVX color
selection (the actual SDSS color selection is done in 3-D color space
and does not use this cut explicitly).  To the limiting magnitude of
the low-$z$ SDSS quasar survey ($i^*\sim19$), the shifting of the
stellar locus as a function of metallicity should have little effect
upon the efficiency of quasar target selection at $z\le2.2$.

Futhermore, it is important to realize that the limiting magnitude of
a sample is a function of the color of the sources.  For example, note
the decrease in faint, red sources in Figure~\ref{fig:fig14n}.  This
dearth is a result of the fact that objects selected to $g^*=22$, must
be bluer than $u^*$-$g^*=0.3$ in order to be brighter than the $u^*$
``plate limit'' of $u^*\sim22.3$.  Thus a sample that includes objects
as red as $u^*$-$g^*=0.8$ can only be considered to be complete to
$g^*=21.5$.

\subsection{Extended Objects\label{sec:seyferts}}

One of the many interesting questions that we can address with the
SDSS quasar sample is whether quasars and Seyferts form a continuum,
or whether they represent physically distinct classes of objects.  To
address this issue it is necessary to target both point sources and
extended sources when looking for quasars; the SDSS quasar target
selection algorithm will not explicitly distinguish between quasars
and Seyferts.  However, it is equally important that we not target too
many normal galaxies during the process of looking for quasars.  As
such, we must understand what regions of color space are populated by
quasars with unresolved image profiles.

In our sample, extended source quasars are typically redder than point
source quasars by $\sim 0.2$ in both $u^*$-$g^*$ and $g^*$-$r^*$.  In
$r^*$-$i^*$ and $i^*$-$z^*$ the colors of extended and point sources
are more similar.  That this is the case can be seen in
Figure~\ref{fig:fig5n}, where we plot as magenta points those
$z\le0.6$ ``quasars'' that are flagged as extended objects in the SDSS
database.  The discrepancy in colors could have a number of origins,
the most likely is that the extended sources are redder in the blue
colors because of contamination from starlight from the quasar host
galaxies.  Such an effect could also be due to the strength of the
small blue bump in low-redshift AGN; we intend to investigate this
possibility in future spectroscopic samples.  Another possibility is
that high-luminosity AGN are more able to blow away dust in their
vicinity and are therefore not as reddened.  The SDSS sample will
eventually cover enough area to assemble a large sample of
high-luminosity low-redshift quasars, for comparison with the far more
numerous low-luminosity low-redshift quasars.  This will allow us to
directly test the hypotheses of host galaxy contamination and
luminosity- or redshift-dependent reddening.

We further investigate these possibilities by studying the colors of
the objects in our sample as a function of absolute $g^*$ magnitude.
Seyferts are typically defined as AGN that are fainter than $M_B =
-23$ and have broad emission lines.  Whether or not this definition
represents a distinction between two physically separate classes of
objects is not at all clear \citep{sg83}.  Furthermore, Seyfert 2s,
like Seyfert 1s have $M_B>-23$, but relatively narrow emission lines.
It is an open question whether a population of ``Quasar 2s'' exists
with space density relative to normal quasars comparable to the
relative space density of Seyfert 2s and Seyfert 1s.  If not, this may
represent a physical, luminosity-dependent difference between these
classes of objects.  Throughout, we refer to Seyfert 1s simply as
Seyfert galaxies.

The problem with using $M_B = -23$ as the dividing line between
quasars and Seyferts is that in any single survey with an optical flux
limit, it is predominantly a division in redshift.  At high redshift,
the limiting magnitude of a sample causes a selection effect such that
only quasars that are intrinsically bright are found.  At low
redshift, intrinsically faint objects are bright enough to make it
into the sample; however, the volume of space sampled is so small that
very few intrinsically bright objects are found.  A selection
criterion based on absolute magnitude is therefore essentially a
redshift cut.

We test to see if the traditional dividing line between quasars and
Seyferts is appropriate, and if so, physical.  If there exists a
correlation between the absolute magnitude and the median-corrected
colors, then it might be possible to use colors to discriminate
quasars from Seyferts.  We ask if there is any such evidence that the
colors of quasars are a function of absolute magnitude.

The average absolute magnitude of quasars in our sample is a steep
function of redshift for $z<1$, so we cannot simply ask if the colors
of all $z<1$ quasars are a function of absolute magnitude.  However,
if we limit ourselves to small regions of redshift space, then we can
assume that the average absolute magnitude is not changing across the
(small) redshift bin.  We have broken our low-redshift data set into
redshift slices of $\Delta z = 0.1$ from $z=0.1$ to $z=0.6$.  We find
that within these redshift slices there is a correlation between the
color and the absolute magnitude.

Since the average $M_{g^*}$ changes significantly from bin to bin, we
cannot simply make a plot of $M_{g^*}$ versus color for the entire
redshift range.  Instead we have chosen to subtract the average
$M_{g^*}$ in each redshift bin and then stack the results for each of
the six redshift slices.  The result is Figure~\ref{fig:fig15n}.  Here
we plot the {\em residual} color versus the {\em normalized}
absolute magnitude for objects with $0.1 \le z \le 0.6$.  If the
colors of these objects were {\em not} a function of absolute
magnitude, then the colors would be scattered around a color of zero.
Instead we see that there is a strong correlation as a function of
absolute magnitude in the sense that brighter objects (more negative
normalized absolute magnitude) tend to be bluer.  Furthermore, the
redder objects have a tendency to be extended sources as might be
expected if the fainter sources have a larger fraction of their light
coming from the stars of the host galaxy.  No such color-magnitude
correlation is seen for a control sample of higher luminosity objects
at higher redshift, again as expected if host galaxy contamination is
the cause.

We conclude the following from Figure~\ref{fig:fig15n}.  Quasars and
Seyfert galaxies form a continuum of properties.  There is no apparent
break in the color properties of quasars as function of absolute
magnitude that would suggest a real physical difference between
quasars and Seyfert galaxies. However, we do find that a cut in
absolute magnitude at or near $M_{g^*}\approx-23$ does exclude most of
the extended sources from the quasar sample and that extended sources
are significantly redder than point sources (in the blue colors).
This distinction might be due to a physical difference in the strength
of the 3000A bump between Seyferts and quasars, or due to the fact
that the former have a larger contribution from their host galaxies.
Arguing against the latter is the fact that the correlation between
color and absolute magnitude holds for unresolved objects as well.
More work is required to fully understand Figure~\ref{fig:fig15n} and
what it says about the differences between low- and high-luminosity
AGN.

\section{Conclusions}

We have presented an empirical investigation of the colors of 2625
quasars ($0.04 \le z \le 5.28$) in the SDSS photometric system.  The
number of quasars with accurate photometry that will result from the
Sloan Digital Sky Survey will spur significant advances in quasar
science in the coming years.  The quality of the photometry allows for
investigations that have hitherto been reserved for spectroscopic
analysis.  The combination of SDSS photometry and spectroscopy
together will enable the resolution of a number of issues that
currently perplex the quasar community.  We have addressed some of
those issues herein and have drawn the following conclusions.

Quasar colors vary significantly as a function of redshift, yet at a
given redshift, the distribution of colors is surprisingly narrow.  In
four dimensional color space, the colors of quasars are largely
non-degenerate as a function of redshift.  As a result, it may be
possible to determine photometric redshifts for quasars, even at low
redshifts.

Typical colors of quasars are consistent with a distribution of
spectral indices centered on $\alpha = -0.5$ with a spread of
$\pm0.65$ (95\% confidence).  We find that the scatter in the colors
of quasars at a given redshift must largely be due to a range of
spectral indices as opposed to a range of emission line strengths,
since the scatter is mostly independent of redshift for $z>0.5$.
However, strong emission lines such as Lyman-$\alpha$, \ion{C}{4},
\ion{Mg}{2}, H$\beta$, H$\alpha$, and even the $\lambda\,3000$ bump
can have a significant affect upon the broad-band colors of quasars.
At low redshift, there is a larger scatter in the colors of quasars in
the blue colors.  This scatter may result from a wider range of
strengths of the $\lambda\,3000$ bump in low-redshift, low-luminosity
quasars, and/or the low-redshift sample may be significantly
contaminated by stellar light from the host galaxies.

Our sample shows no significant break in the colors of lower
luminosity quasars (Seyfert galaxies) as compared to higher luminosity
objects.  However, we do find that lower luminosity quasars are redder
than their higher luminosity counterparts for $0.05 < z < 0.65$.  We
find that the effect of using the traditional dividing line of $M_B =
-23$ to separate the populations is to exclude all of the lowest
redshift AGN and most of those that are contaminated by stellar light
from the host galaxy.  However, it is unclear if this division is
physical or simply observational.

Finally, we have discovered a statistical sample of reddened quasars.
We find that the number of quasars with colors appreciably redder than
the median is larger than might be expected in the absence of
reddening.  This reddening is probably not the result of stellar light
from the host galaxy or absorption from intervening galaxies, since
the reddening is not a function of redshift.  Instead the observed
reddening is probably internal to the quasars.  Further examination of
these quasars on an individual and on a group basis is in progress.

\acknowledgements

The Sloan Digital Sky Survey\footnote{The SDSS Web site is
http://www.sdss.org/.} (SDSS) is a joint project of The University of
Chicago, Fermilab, the Institute for Advanced Study, the Japan
Participation Group, The Johns Hopkins University, the
Max-Planck-Institute for Astronomy, New Mexico State University,
Princeton University, the United States Naval Observatory, and the
University of Washington. Apache Point Observatory, site of the SDSS,
is operated by the Astrophysical Research Consortium (ARC).  Funding
for the project has been provided by the Alfred P. Sloan Foundation,
the SDSS member institutions, the National Aeronautics and Space
Administration, the National Science Foundation, the U.S. Department
of Energy, Monbusho, and the Max Planck Society.  The Hobby-Eberly
Telescope (HET) is a joint project of the University of Texas at
Austin, the Pennsylvania State University, Stanford University,
Ludwig-Maximillians-Universit\"at M\"unchen, and
Georg-August-Universit\"at G\"ottingen.  The HET is named in honor of
its principal benefactors, William P. Hobby and Robert E. Eberly.
This research has made use of the NASA/IPAC Extragalactic Database
(NED) which is operated by the Jet Propulsion Laboratory, California
Institute of Technology, under contract with the National Aeronautics
and Space Administration.  DPS and GTR acknowledge support from NSF
grant AST99-00703.  XF and MAS acknowledge support from Research
Corporation, NSF grants AST96-16901 and AST-0071091, the Princeton
University Research Board, and a Porter O. Jacobus Fellowship.  We
thank Sofia Kirhakos and Emily Leibacher for their assistance in an
early stage of this project.  We thank Bob Becker for providing data
in advance of publication.  We thank Pat Hall and David Weinberg for
critical readings of the manuscript and for the resulting
improvements.

\clearpage

\begin{figure}[p]
\epsscale{1.0}
\plotone{RichardsGT.fig6n.ps}
\caption{Simulated color versus redshift.  The solid black line is the
observed median color-redshift relation from Table~\ref{tab:tab3}.
The dashed black line is the color for a power-law spectrum of the
form $f_{\nu} \propto \nu^{-0.5}$.  The solid blue line is
color-redshift track of the SDSS quasar composite spectrum
\citep{vbr00,van+01}. The remaining lines show the colors as a
function of redshift when specific emission features are removed from
the composite spectrum.\label{fig:fig6n}}
\end{figure}

\begin{figure}[p]
\epsscale{1.0}
\plotone{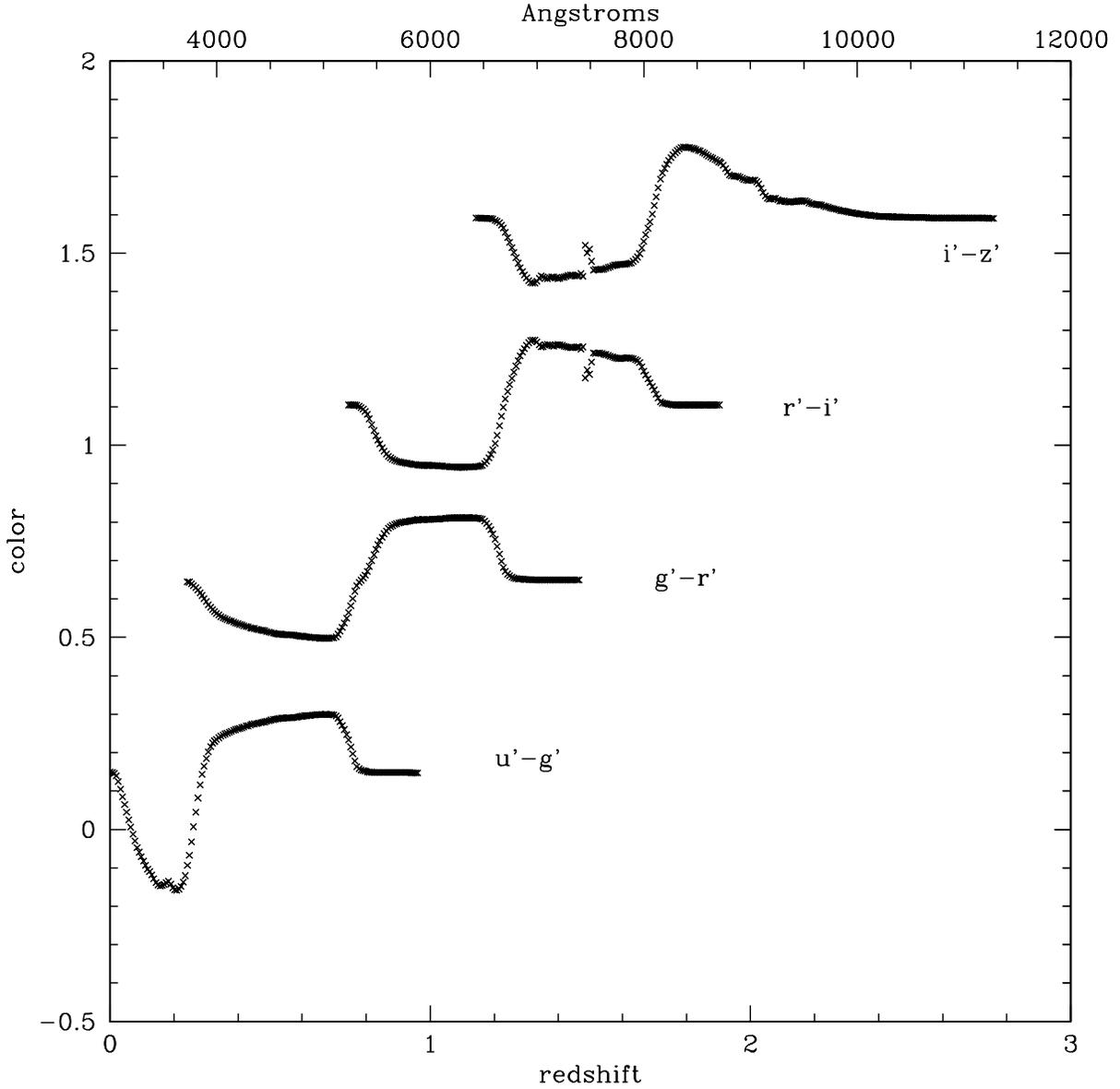}
\caption{Color versus redshift for a top hat emission line
superimposed upon a power-law continuum ($\alpha_{\nu}=-0.5$).  The
line center of the top hat begins at $3000\,{\rm \AA}$ and is
redshifted through each of the transmission curves.  The lower axis
gives the redshift of the line, whereas the upper axis gives the
wavelength of the line.  The top hat has an observed equivalent width
of $200\,{\rm \AA}$ and is $20\,{\rm \AA}$ wide.  The bottom,
left-hand curve is the $u'-g'$ curve.  The other colors are shifted in
the y-direction by 0.5 each.\label{fig:fig7n}}
\end{figure}

\begin{figure}[p]
\epsscale{1.0}
\plotone{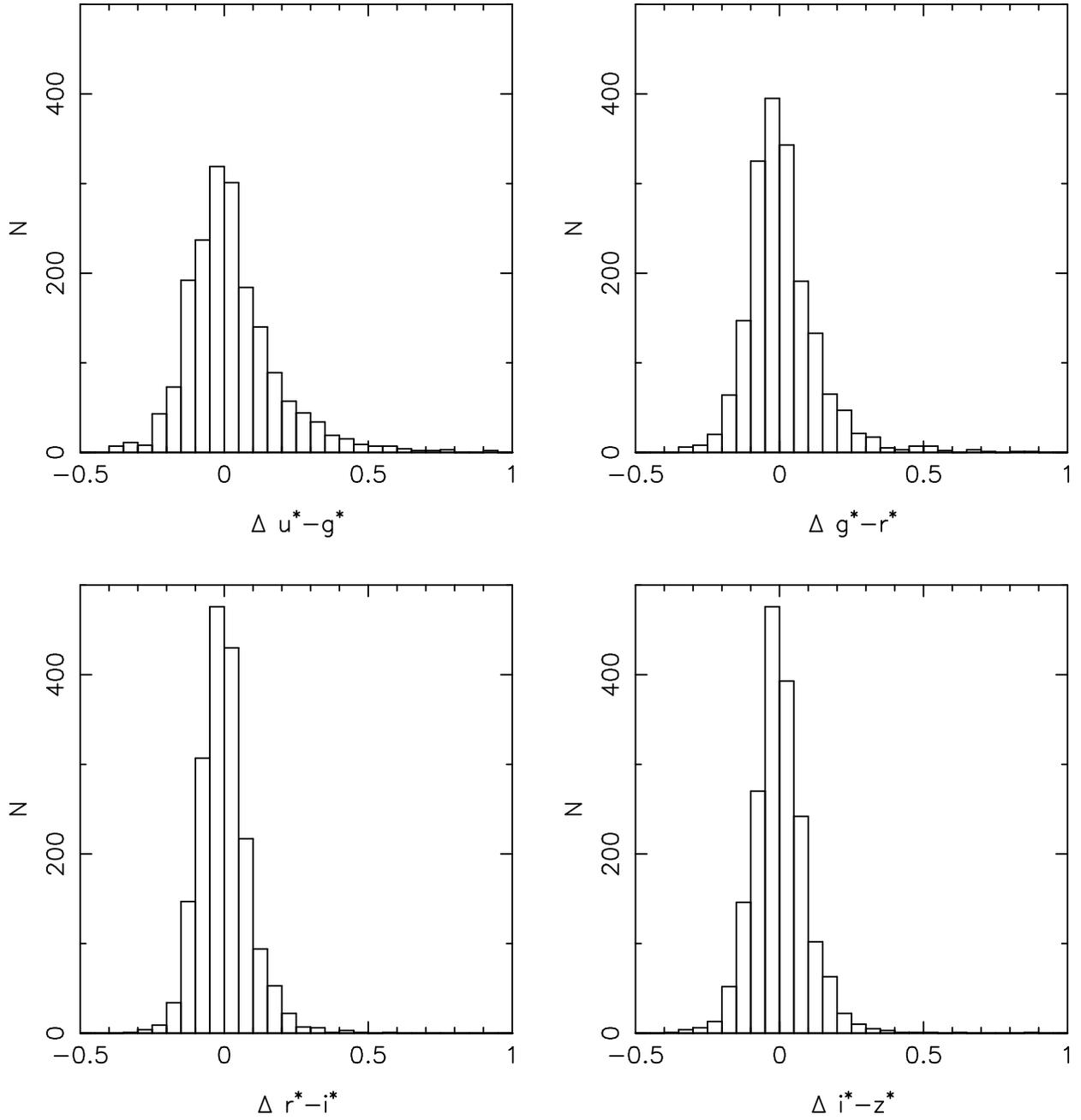}
\caption{Histograms of observed quasar colors corrected by the median
colors as a function of redshift.\label{fig:fig8n}}
\end{figure}

\begin{figure}[p]
\epsscale{1.0}
\plotone{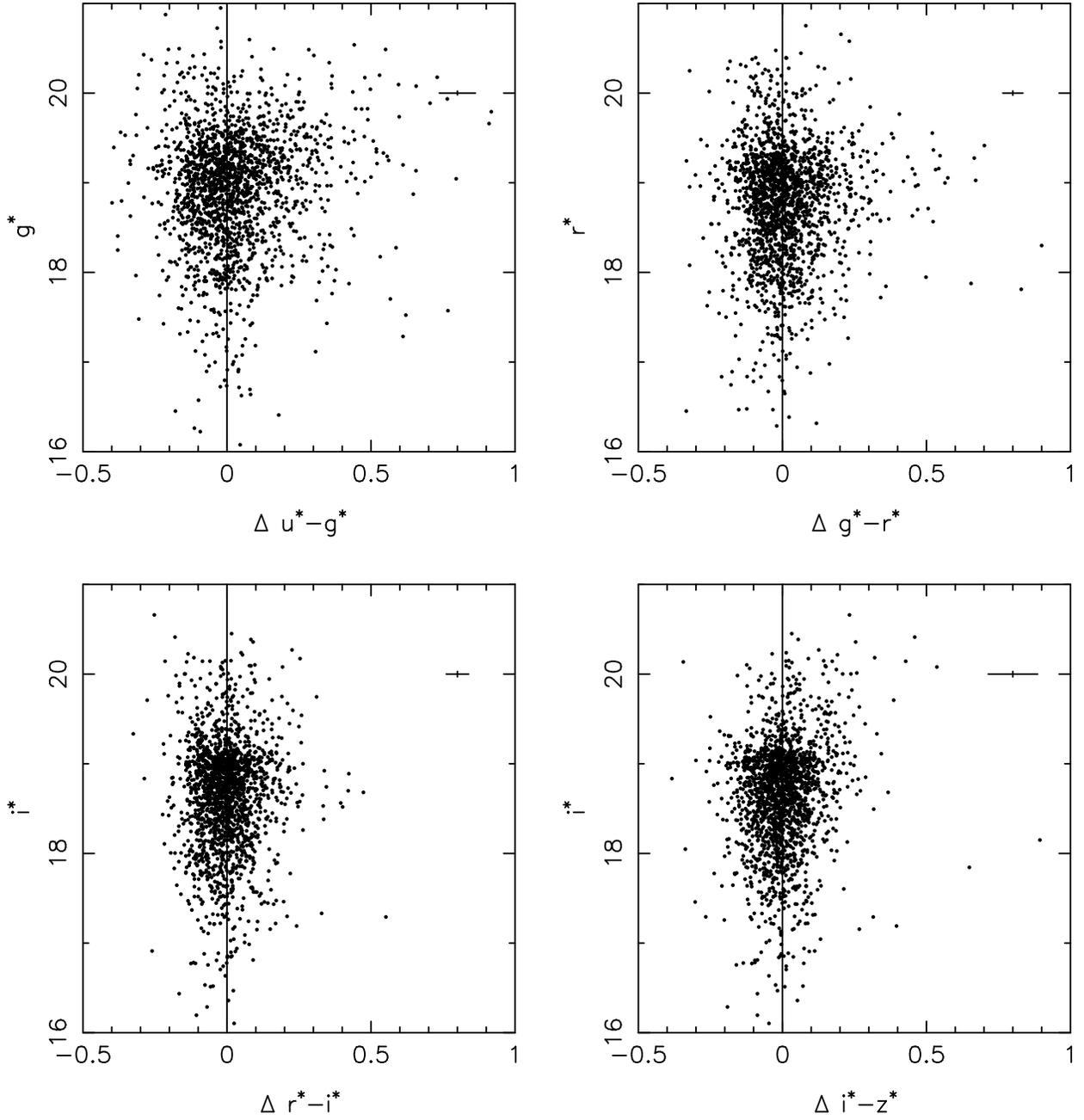}
\caption{Observed quasar colors corrected by the median colors as a
function of redshift, plotted as a function of magnitude.  Two-sigma
error bars in color and magnitude are given for the average $20^{\rm
th}$ magnitude object.\label{fig:fig9n}}
\end{figure}

\begin{figure}[p]
\epsscale{1.0}
\plotone{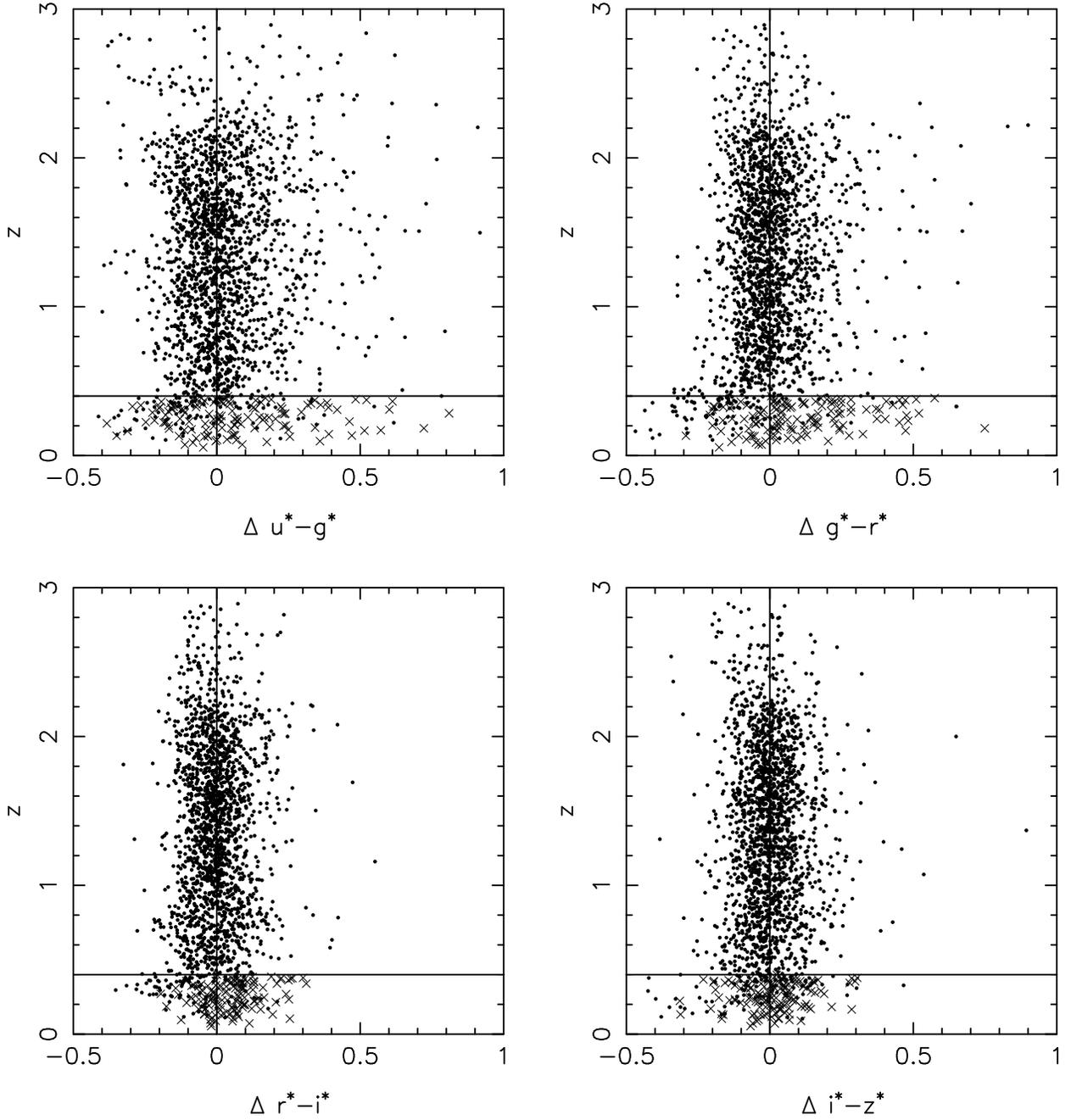}
\caption{Quasar colors (corrected by the median colors as a function
of redshift), plotted versus redshift.  Objects with $z\le0.4$
(horizontal line) are not used in the reddening analysis.  Crosses
mark extended sources with $z\le0.4$.\label{fig:fig10n}}
\end{figure}

\begin{figure}[p]
\epsscale{1.0}
\plotone{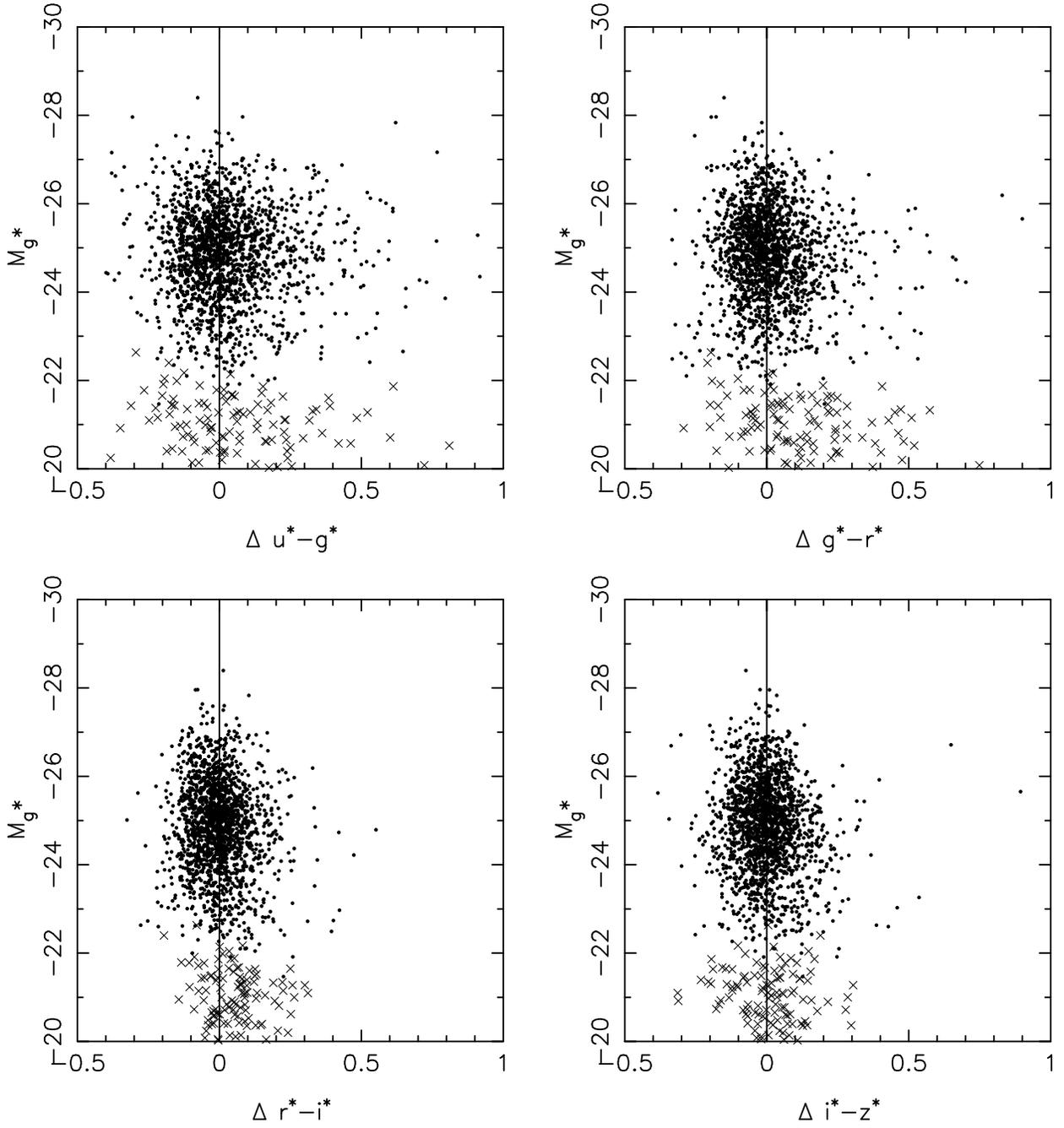}
\caption{Quasar colors (corrected by the median colors as a function
of redshift), plotted versus absolute magnitude.  Extended sources
with $z\le0.4$ are plotted as crosses.\label{fig:fig11n}}
\end{figure}

\begin{figure}[p]
\epsscale{1.0}
\plotone{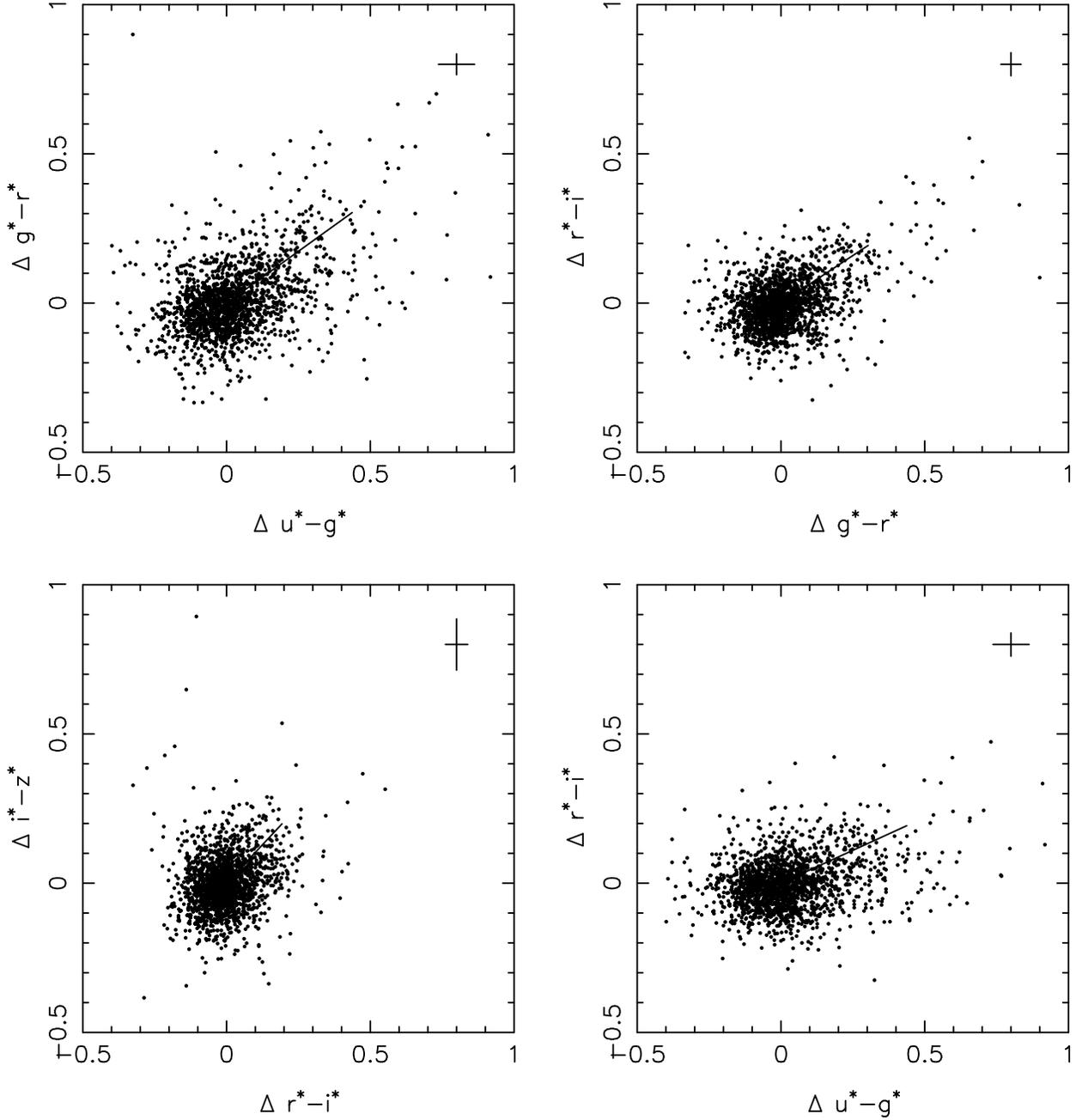}
\caption{Quasar colors (corrected by the median colors as a function
of redshift), plotted against each other.  Typical errors are given in
the upper right hand corner of each panel.  The vector in each panel
is the absolute value of the blue 95\% confidence limit from
Table~\ref{tab:tab5}.
\label{fig:fig12n}}
\end{figure}

\begin{figure}[p]
\epsscale{1.0}
\plotone{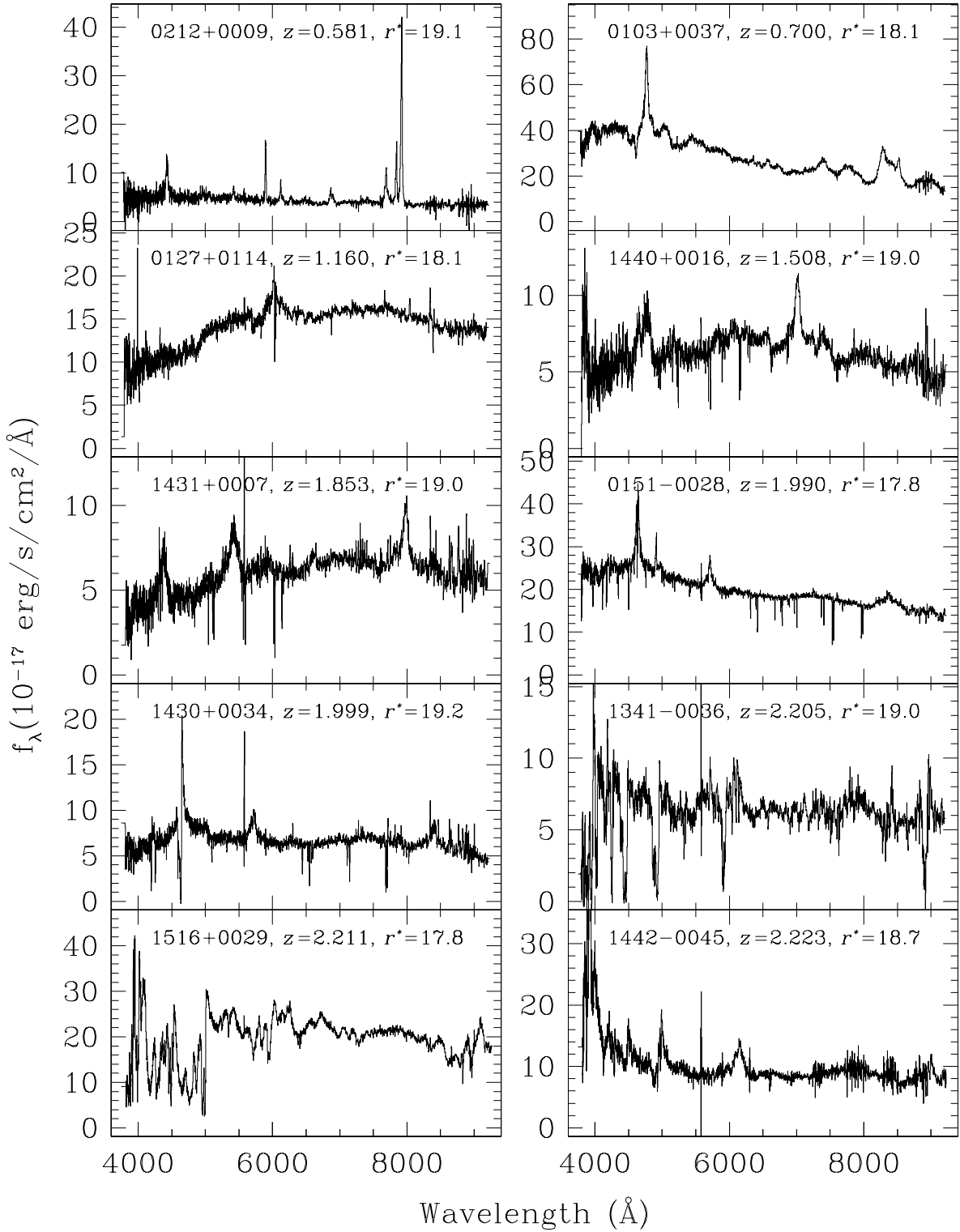}
\caption{Sample ``red'' quasars (SDSSp J\#\#\#\#$\pm$\#\#\#\#).  That these
quasars are anomalously red can be seen by comparing them to the more
``normal'' quasars at the same redshift shown in
Figure~\ref{fig:fig1n}.\label{fig:fig13n}}
\end{figure}

\begin{figure}[p]
\epsscale{1.0}
\plotone{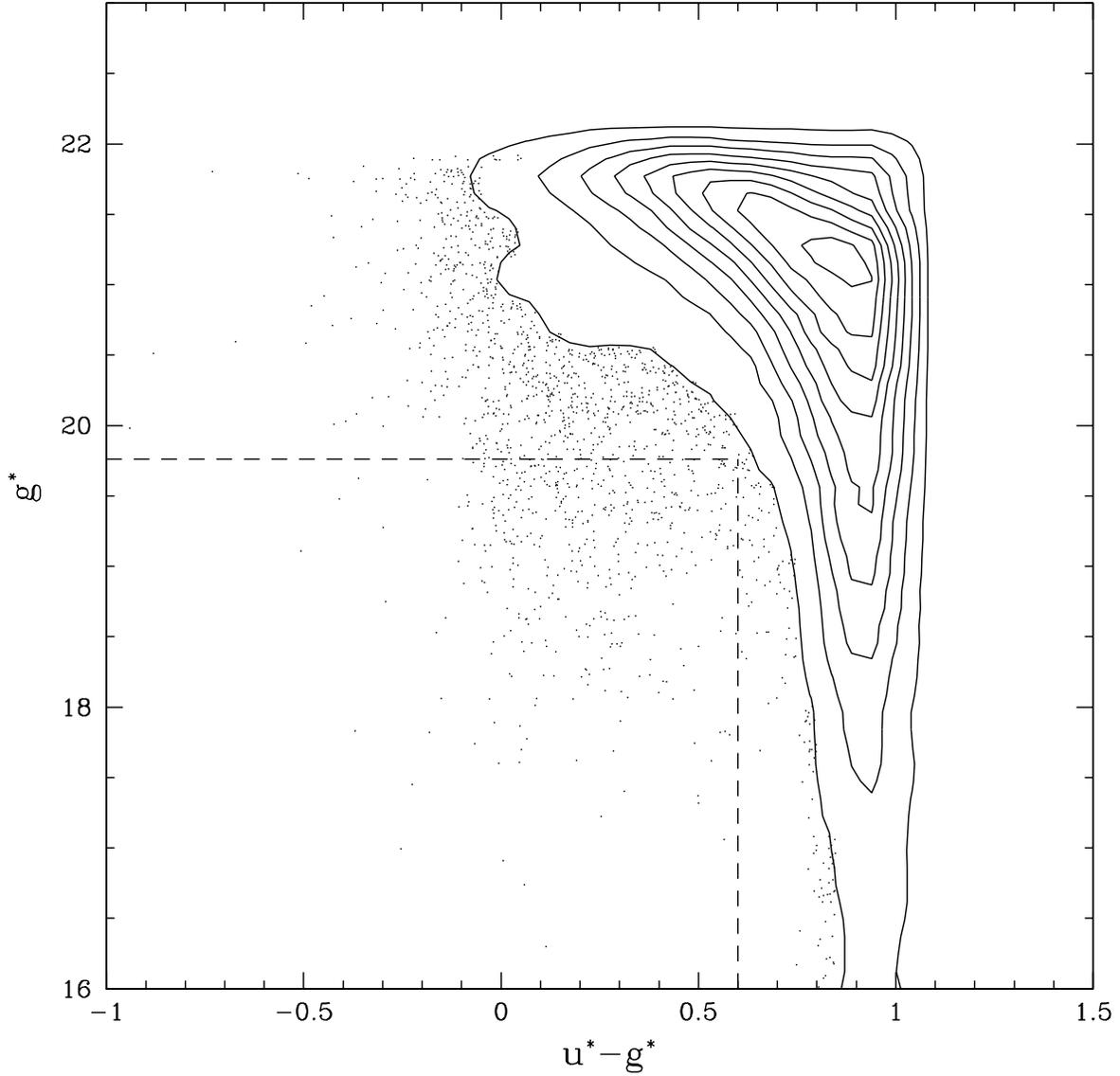}
\caption{Stellar UVX Sources. Stellar objects from camera column 3 of
run 756 with $u^*$-$g^*\le1.0$ and $g^*\le22.0$.  The steep fall-off
of the red sources is artifical: it results from the red color-cut on
the data.  The vertical dashed lines shows an appropriate color-cut
for UVX quasars in the SDSS system.  The horizontal dashed line shows
the average $g^*$ magnitude for a low-redshift quasar with $i^*=19.0$.
Note the shifting of the stellar locus as a function of magnitude and
the fall-off of faint, red sources.\label{fig:fig14n}}
\end{figure}

\begin{figure}[p]
\epsscale{1.0}
\plotone{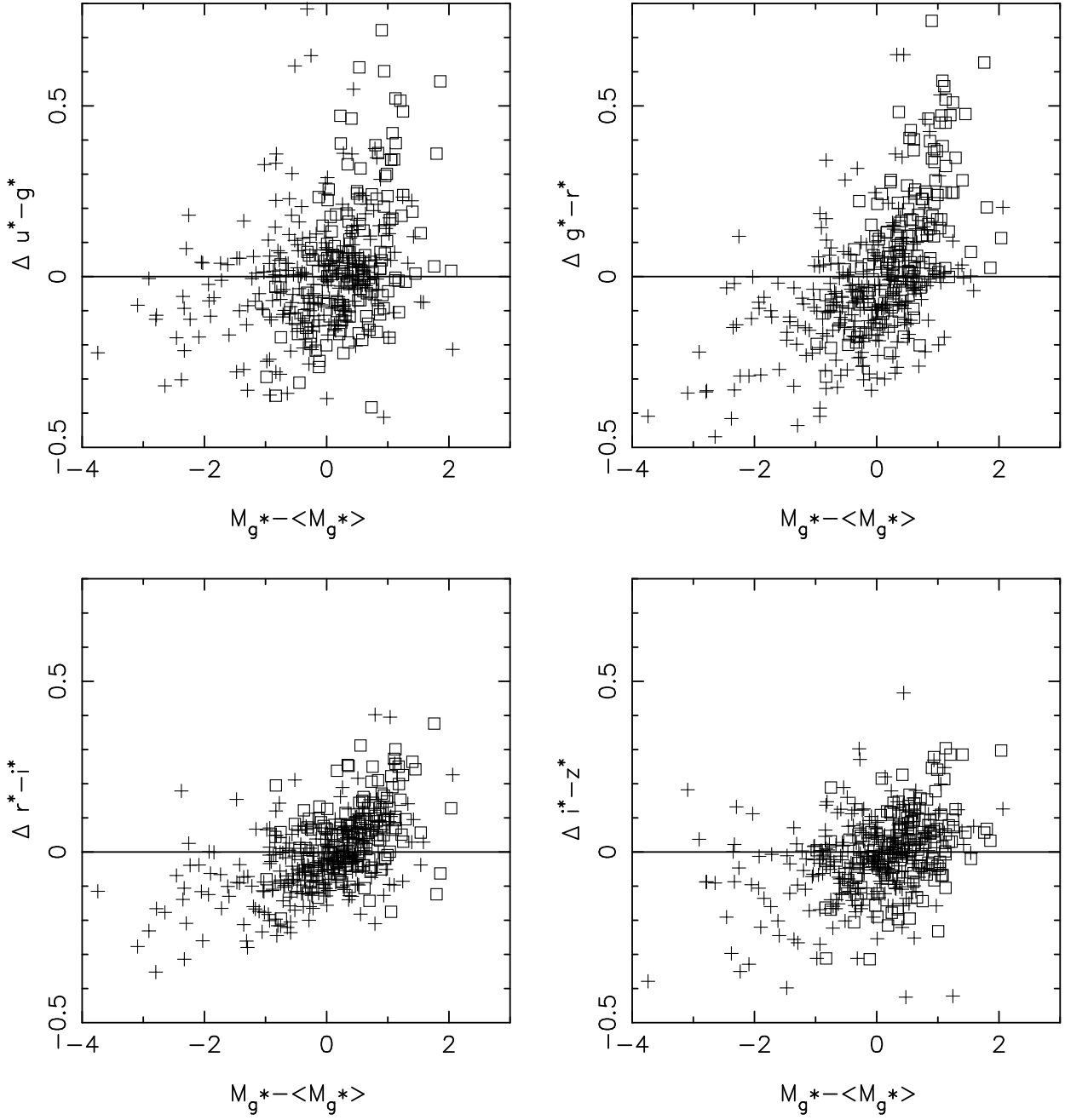}
\caption{Quasar colors (corrected by the median colors as a function
of redshift), plotted versus normalized absolute magnitude.  The
absolute magnitude of the objects is normalized in redshift bins of
$\Delta z = 0.1$ from $z=0.1$ to $z=0.6$.  Crosses are point sources,
whereas squares represent extended sources.  The lines show the median
corrected color around which the points would be evenly distributed if
there were not a correlation between color and luminosity.  The mean
absolute magnitude ranges from $M_{g^*} = -20.1$ in the $z=0.1$ bin to
$M_{g^*} = -23.5$ in the $z=0.6$ bin.\label{fig:fig15n}}
\end{figure}

\clearpage

\begin{deluxetable}{rlllc}
\tablewidth{0pt}
\tablecaption{Summary of Observations \label{tab:tab1}}
\tablehead{
\colhead{Run} & 
\colhead{UT Date} & 
\colhead{Strip} & 
\colhead{RA Range} &
\colhead{Seeing} \\
\colhead{} & 
\colhead{} & 
\colhead{} & 
\colhead{(J2000)} &
\colhead{($\arcsec$)} 
}
\startdata
94 & 1998 Sep 19 & North & 22:16 - 03:49 & 1.5 \\
125 & 1998 Sep 25 & South & 23:18 - 05:17 & 1.5 \\
752 & 1999 Mar 21 & South & 08:21 - 15:41 & 1.5-1.8 \\
756 & 1999 Mar 22 & North & 08:21 - 16:34 & 1.2-1.5 \\
\enddata
\end{deluxetable}

\begin{deluxetable}{llllllllllcl}
\tabletypesize{\tiny}
\rotate
\tablewidth{0pt}
\tablecaption{SDSS Quasar Photometry \label{tab:tab2}}
\tablehead{
\colhead{Name} &
\colhead{$z_{em}$} &
\colhead{RA(2000)} &
\colhead{DEC(2000)} &
\colhead{$\Delta \arcsec$} &
\colhead{$u^{*}$} &
\colhead{$g^{*}$} &
\colhead{$r^{*}$} &
\colhead{$i^{*}$} &
\colhead{$z^{*}$} &
\colhead{E($g^{*}$-$r^{*}$)} &
\colhead{source}}
\startdata
LBQS 2357-0014 & 0.479 & $ 00:00:11.96 $ & $ +00:02:25.3 $ & 0.11 & 17.83$\pm$0.011 & 17.58$\pm$0.01 & 17.70$\pm$0.01 & 17.65$\pm$0.01 & 17.66$\pm$0.024 & 0.03 & 1 2\\ 
LBQS 2358+0038 & 0.954 & $ 00:00:42.91 $ & $ +00:55:39.7 $ & 0.25 & 18.34$\pm$0.016 & 18.19$\pm$0.01 & 17.93$\pm$0.01 & 17.88$\pm$0.01 & 17.69$\pm$0.023 & 0.03 & 1 2\\ 
ZC 2358-010B & 1.890 & $ 00:00:58.24 $ & $ -00:46:46.3 $ & 0.39 & 18.55$\pm$0.018 & 18.68$\pm$0.01 & 18.71$\pm$0.011 & 18.40$\pm$0.010 & 18.32$\pm$0.036 & 0.04 & 1 2\\ 
UM 194 & 0.780 & $ 00:01:16.53 $ & $ +00:02:05.4 $ & 1.67 & 18.94$\pm$0.025 & 18.79$\pm$0.01 & 18.94$\pm$0.012 & 19.04$\pm$0.017 & 19.18$\pm$0.087 & 0.03 & 1\phm{ 2}\\ 
\vspace{0.03in}
{[HB89]} 2359-000 & 2.120 & $ 00:01:46.95 $ & $ +00:14:28.9 $ & 2.03 & 20.92$\pm$0.087 & 20.70$\pm$0.029 & 20.66$\pm$0.038 & 20.53$\pm$0.057 & 20.14$\pm$0.189 & 0.03 & 1\phm{ 2}\\ 
{[HB89]} 2359+002 & 2.670 & $ 00:02:12.01 $ & $ +00:32:53.5 $ & 2.47 & 20.50$\pm$0.084 & 19.48$\pm$0.015 & 19.42$\pm$0.018 & 19.31$\pm$0.026 & 19.26$\pm$0.081 & 0.03 & 1\phm{ 2}\\ 
LBQS 2359-0021 & 0.812 & $ 00:02:22.48 $ & $ -00:04:43.4 $ & 0.13 & 18.84$\pm$0.017 & 18.39$\pm$0.01 & 18.19$\pm$0.01 & 18.16$\pm$0.01 & 17.86$\pm$0.023 & 0.03 & 1 2\\ 
CFRS 00.0207 & 1.352 & $ 00:02:24.50 $ & $ -00:41:28.7 $ & 0.73 & 24.05$\pm$1.127 & 23.10$\pm$0.341 & 22.14$\pm$0.209 & 22.49$\pm$0.423 & 20.68$\pm$0.298 & 0.03 & 1\phm{ 2}\\
\enddata
\tablecomments{Object sources listed in the last column are NED (1) and SDSS (2).}
\end{deluxetable}

\begin{deluxetable}{rcrrrrcrrrrcrrrrcrrrr}
\tabletypesize{\scriptsize}
\rotate
\tablewidth{0pt}
\tablecaption{Color-Redshift Relation \label{tab:tab3}}
\tablehead{
\colhead{} &
\colhead{} &
\multicolumn{4}{c}{$u^*-g^*$} &
\colhead{} &
\multicolumn{4}{c}{$g^*-r^*$} &
\colhead{} &
\multicolumn{4}{c}{$r^*-i^*$} &
\colhead{} &
\multicolumn{4}{c}{$i^*-z^*$} \\
\cline{3-6} \cline{8-11} \cline{13-16} \cline{18-21} \\
\colhead{$z$} &
\colhead{} &
\colhead{Median} &
\colhead{Low\tablenotemark{a}} &
\colhead{High\tablenotemark{b}} &
\colhead{N} &
\colhead{} &
\colhead{Median} &
\colhead{Low\tablenotemark{a}} &
\colhead{High\tablenotemark{b}} &
\colhead{N} &
\colhead{} &
\colhead{Median} &
\colhead{Low\tablenotemark{a}} &
\colhead{High\tablenotemark{b}} &
\colhead{N} &
\colhead{} &
\colhead{Median} &
\colhead{Low\tablenotemark{a}} &
\colhead{High\tablenotemark{b}} &
\colhead{N}
}
\startdata
0.05 & & $0.326$ & $0.183$ & $0.380$ & 4 & & $0.282$ & $0.278$ & $0.369$ & 4 & & $0.379$ & $0.340$ & $0.392$ & 4 & & $0.088$ & $0.024$ & $0.126$ & 4 \\
0.10 & & $0.253$ & $-0.002$ & $0.450$ & 11 & & $0.356$ & $0.156$ & $0.564$ & 14 & & $0.453$ & $0.323$ & $0.705$ & 14 & & $0.135$ & $-0.271$ & $0.242$ & 14 \\
0.15 & & $0.159$ & $-0.170$ & $0.756$ & 25 & & $0.358$ & $-0.101$ & $0.681$ & 31 & & $0.410$ & $0.219$ & $0.569$ & 31 & & $0.056$ & $-0.229$ & $0.258$ & 29 \\
0.20 & & $0.225$ & $-0.180$ & $0.814$ & 29 & & $0.397$ & $0.091$ & $0.826$ & 33 & & $0.362$ & $0.158$ & $0.530$ & 33 & & $0.092$ & $-0.151$ & $0.295$ & 33 \\
0.25 & & $0.153$ & $-0.245$ & $0.458$ & 39 & & $0.348$ & $0.038$ & $0.846$ & 42 & & $0.236$ & $-0.023$ & $0.447$ & 42 & & $0.243$ & $-0.017$ & $0.571$ & 42 \\
0.30 & & $0.208$ & $-0.129$ & $0.615$ & 36 & & $0.385$ & $-0.056$ & $0.784$ & 41 & & $0.154$ & $-0.094$ & $0.339$ & 39 & & $0.486$ & $0.241$ & $0.627$ & 39 \\
0.35 & & $0.131$ & $-0.132$ & $0.677$ & 59 & & $0.254$ & $-0.168$ & $0.956$ & 66 & & $0.092$ & $-0.165$ & $0.364$ & 66 & & $0.513$ & $0.295$ & $0.788$ & 63 \\
0.40 & & $0.183$ & $-0.105$ & $0.492$ & 46 & & $0.095$ & $-0.236$ & $0.725$ & 49 & & $0.122$ & $-0.133$ & $0.411$ & 49 & & $0.370$ & $0.011$ & $0.723$ & 48 \\
\enddata
\tablenotetext{a}{Color at the 95\% confidence limit to the blue of the median}
\tablenotetext{b}{Color at the 95\% confidence limit to the red of the median}
\end{deluxetable}

\begin{deluxetable}{rrcccc}
\tablewidth{0pt}
\tablecaption{Power-law Colors of Quasars \label{tab:tab4}}
\tablehead{
\colhead{$\alpha_{\nu}$} & 
\colhead{$\alpha_{\lambda}$} & 
\colhead{$u'-g'$} & 
\colhead{$g'-r'$} &
\colhead{$r'-i'$} & 
\colhead{$i'-z'$}
}
\startdata
$-2.5$ & 0.5 & 0.756 & 0.739 & 0.523 & 0.459 \\
$-2.0$ & 0.0 & 0.602 & 0.594 & 0.419 & 0.367 \\
$-1.5$ & $-0.5$ & 0.450 & 0.447 & 0.315 & 0.275 \\
$-1.0$ & $-1.0$ & 0.299 & 0.299 & 0.210 & 0.184 \\
$-0.5$ & $-1.5$ & 0.149 & 0.150 & 0.105 & 0.092 \\
0.0 & $-2.0$ & 0.000 & 0.000 & 0.000 & 0.000 \\
\enddata
\end{deluxetable}

\begin{deluxetable}{ccccc}
\tablewidth{0pt}
\tablecaption{Scatter in Corrected Colors (95\% Confidence)\label{tab:tab5}}
\tablehead{
\colhead{Color} &
\colhead{$\Delta$ Blue Color} & 
\colhead{$\Delta$ Red Color} & 
\colhead{$\Delta \alpha$ Blue} & 
\colhead{$\Delta \alpha$ Red}
}
\startdata
$u^*$-$g^*$ & $-0.209$ & 0.402 & 0.69 & 1.34 \\
$g^*$-$r^*$ & $-0.190$ & 0.319 & 0.64 & 1.07 \\
$r^*$-$i^*$ & $-0.161$ & 0.198 & 0.77 & 0.94 \\
$i^*$-$z^*$ & $-0.173$ & 0.211 & 0.95 & 1.15 \\
\enddata
\end{deluxetable}

\end{document}